\documentstyle[preprint,eqsecnum,aps]{revtex}

\begin{document}
\draft

\title{Metal-insulator transition in a doubly orbitally degenerate model
with correlated hopping}
\author{L.~Didukh\cite{e-mail}, Yu.~Skorenkyy, Yu.~Dovhopyaty, V.~Hankevych}
\address{Ternopil State Technical University, Department of Physics,\\
56 Rus'ka Str., Ternopil UA--46001, Ukraine }
\date{\today}
\maketitle
\begin{abstract}
In the present paper we propose a doubly orbitally degenerate narrow-band
model with correlated hopping. The peculiarity of the
model is taking into account the matrix element of electron-electron
interaction which describes intersite hoppings of electrons. In particular, 
this leads to the concentration dependence of the effective hopping 
integral. The cases of the strong and weak Hund's coupling are 
considered. By means of a generalized mean-field approximation 
the single-particle Green function and quasiparticle energy 
spectrum are calculated. Metal-insulator transition is studied in the model
at different integer values of the electron concentration. With the help
of the obtained energy spectrum we find energy gap width and criteria of
metal-insulator transition. 
\end{abstract}


\narrowtext

\section{Introduction}
\label{sec: intr}

The electron-electron interaction-driven metal-insulator transition (MIT)
has fascinated theorists and experimentalists for many years.
This transition is named after Sir Nevill Mott, being one of the pioneers 
who laid down the foundations of our physical understanding of 
this phenomenon \cite{1,1_69}.
Despite a large amount of papers devoted to study of this transition 
(e.g. see a review \cite{1_11}) the construction 
of consistent theory of MIT is still far from carrying out and constitutes one 
of the most challenging problems in condensed matter physics.

The simplest model describing MIT in materials with narrow energy
bands is the Hubbard model \cite{1_1}. This model describes a single 
non-degenerate band of electrons with the local Coulomb interaction. The 
model Hamiltonian contains two energy parameters: the matrix element
$t_0$ being the hopping integral of an electron from one site to another
($t_0$ is not dependent on occupation of sites involved in the hopping 
process) and the parameter $U$ being the intraatomic Coulomb repulsion of 
two electrons of the opposite spins. This model is studied intensively (for 
recent reviews see Refs. \cite{1_9,1_10}).

Theoretical analyses, on the one hand, and available experimental data, on 
the other hand, point out the necessity of the Hubbard model generalization.
As a rule this generalization is perfomed by two means: taking into 
account the orbital degeneration or correlated hopping (in the present 
paper we do not consider the generalization of the Hubbard model by 
taking into account the inter-atomic Coulomb and exchange interactions). 

A model of non-degenerate band has to be generalized by
taking into account the orbital degeneration being a characteristic of 
the transition metal compounds which exhibit MIT. 
First the degenerate Hubbard model was introduced for
description of transition metal compounds 
in papers of Roth~\cite{2}, Kugel' and Khomskii~\cite{3}, Cyrot and
Lyon-Caen~\cite{4}. In these works, in particular, the importance
of intra-atomic exchange interaction $J$ which stabilizes the localized
magnetic moments in accordance with the Hund's rule was investigated.

The intensive study of MIT in the degenerate Hubbard model
has begun only in few recent years by use of the slave boson 
method~\cite{5,12,6,7},
the variational method~\cite{8}, the limit of infinite dimension~\cite{9}.

The results of works~\cite{5,12,6,7,8,9} show the
possibility of the
metal-insulator transition in the doubly degenerate Hubbard model.
However the criteria of the transition obtained in the noted works
are substantially different. 
In particular, at $n=1,\ J=0$ (here in all the
cases $T=0$~K and the rectangular density of states are considered)
in Refs.~\cite{12,6} the criterion of MIT is $U/w=4.95$, 
in Ref.~\cite{8} -- $U/w=5.18$,
in Ref.~\cite{9} -- $U/w=3.0$;
at $n=1,\ J=0.1$ in paper~\cite{5} -- $U/w=3.04$;
at $n=2,\ J=0$
in Refs.~\cite{12,6} -- $U/w=6.0$,  
in Ref.~\cite{5} $U/w=4.0$, in Ref.~\cite{8} -- $U/w=9.0$,
in Ref.~\cite{9} -- $U/w=3.7$.
In addition in paper \cite{5}  at $n=1$ MIT of first order 
is found,
but in papers~\cite{6} and~\cite{8} -- 
MIT of second order.
Consequently a further investigation of MIT in narrow-band models with 
orbital degeneration is necessary. 

Another way of the Hubbard model generalization which allows
to describe the essential peculiarities of transition metal compounds
is taking into account the correlated hopping.
The necessity of taking into consideration correlated hopping is 
caused by two reasons. Firstly, theoretical 
analysis \cite{1_14} points out the inapplicability of the Hubbard model 
for the description of real strongly correlated electron systems, in some 
compounds (e.g. see the estimation in 
Refs.~\cite{1_12,1_38,1_39,1_40,1_42}) the
matrix element of electron-electron interaction describing correlated 
hopping is the same order that the hopping integral or on-site Coulomb 
repulsion. Secondly, using the idea of correlated hopping and caused by it 
the electron-hole asymmetry we can interpret the peculiarities of some
physical properties of narrow-band 
materials \cite{1_13,1_31,1_32,1_33,1_25,1_6,1_26}.

Now two ways are commonly used to generalize the single-band Hubbard model
by taking into account correlated hopping. One of them has been proposed in
Ref. \cite{1_14}. Hirsch showed that in contrast to the hopping integral of
the Hubbard model (which is not dependent on occupation of sites involved in
the hopping process) this integral of a generalized Hubbard model had to
depend on occupation of sites involved in the hopping process.
Hamiltonian of the generalized in such a way Hubbard model is written  as
\begin{eqnarray} \label{G_ham}
&&H=-\sum \limits_{ij\sigma} t_{ij}^{\sigma}a_{i\sigma}^{+}a_{j\sigma}+
U \sum_{i}n_{i\uparrow}n_{i\downarrow},
\\ \label{h_i_G_ham}
&& t_{ij}^{\sigma}=t_{AA}(1-n_{i{\bar \sigma}})(1-n_{j{\bar \sigma}})+
t_{AB}(n_{i{\bar \sigma}}+n_{j{\bar \sigma}}-2n_{i{\bar \sigma}} 
n_{j{\bar \sigma}})+t_{BB}n_{i{\bar \sigma}}n_{j{\bar \sigma}}.
\end{eqnarray}

In recent few years Hamiltonian (\ref{G_ham}) is widely used to study
MIT in narrow energy bands \cite{1_16,1_17,1_20}.

In Ref. \cite{1_12,1_6} the necessity of the Hubbard model generalization
by taking into account the matrix element of electron-electron interaction
describing intersite hoppings of electrons had been pointed out.
The Hamiltonian of the generalized Hubbard model with correlated hopping is
\begin{eqnarray} \label{ham}
H=-\mu \sum_{i\sigma}a_{i\sigma}^{+}a_{i\sigma}+
t(n){\sum \limits_{ij\sigma}}'a_{i\sigma}^{+}a_{j\sigma}+
X{\sum \limits_{ij\sigma}}' \left(a_{i\sigma}^{+}a_{j\sigma}n_{i\bar \sigma}
+h.c.\right)
+U \sum_{i}n_{i\uparrow}n_{i\downarrow},
\end{eqnarray}
with
\begin{eqnarray} \label{c_d_h}
t(n)=t_0+n\sum_{\stackrel{k\neq{i}}{k\neq{j}}}J(ikjk)
\end{eqnarray}
being the effective hopping integral of electrons between nearest-neighbor 
sites of lattice,
$X=J(iiij);\ n=N_e/N$ is the electron concentration ($N_e$ is the number of
electrons, $N$ is the number of lattice sites), 
\begin{eqnarray}
J(ikjk)=\int \int \varphi^*({\bf r}-{\bf R}_i)\varphi({\bf r}-{\bf R}_j)
{e^2\over |{\bf r}-{\bf r'}|}|\varphi({\bf r'}-{\bf R}_k)|^2{\bf drdr'},
\end{eqnarray}
$\varphi({\bf r}-{\bf R}_i)$ is the Wannier function, 
the prime at the sums signifies that $i\neq j$.

In the model described by Hamiltonian~(\ref{ham}) an electron hopping from 
one site to another is correlated both by the occupation of the sites 
involved 
in the hopping process (with the hopping integral $X$) and the occupation of
the nearest-neighbor sites (with the matrix element $J(ikjk)$ at 
$k\ne i,\ k\ne j$) which we take into
account in the Hartree-Fock approximation (Eq.~(\ref{c_d_h})).
The peculiarity of model~(\ref{ham}) is the concentration dependence 
of the hopping integral $t(n)$ in contrast to similar models.

MIT in a generalized Hubbard model with correlated hopping
has been studied in a number of recent 
works \cite{1_16,1_17,1_20,1_63,1_64,1_65,1_59,1_61,1_62,1_51}.
In particular, at half-filling and $t_0=-X$ (or $t_{AB}=0$) some exact
results have been found \cite{1_16,1_63,1_64,1_65}. In a simple cubic
lattice with coordination number $z$ MIT occurs at
\begin{eqnarray} \label{exact_cr}
U_c=z(|t_{AA}+t_{BB}|)=2z|t_0|.
\end{eqnarray}   
If $U>U_c$ the ground state of system is a paramagnetic Mott-Hubbard 
insulator with the concentration of doubly occupied sites $d=0$, the ground
state energy is equal to zero.

For an arbitrary $t_0\ne -X$ (or $t_{AB}\ne 0$) the finding of MIT criterion  
and the description of this phenomenon in a generalized Hubbard model with
correlated hopping still remain an open problem. One of the step to solve
this task is recent papers \cite{1_17,1_20,1_59,1_61,1_62,1_51} where
criteria of MIT, ground state energy, concentration of doubly occupied 
sites have been found. In Refs. \cite{1_17,1_20,1_59,1_61,1_62} the
authors have obtained the following criterion of MIT:
\begin{eqnarray}
U_c=z(|t_{AA}|+|t_{BB}|)=z(|t_0|+|t_0+2X|)
\end{eqnarray}
in agreement with the Mott's general physical ideas \cite{1_69}.
By means of the slave bosons method \cite{1_53} it has been found in 
Ref. \cite{1_51} that MIT occurs at $U_c=4z|t+X|$; however here there 
is a problem of
discrepancy of this result with the exact MIT criterion (\ref{exact_cr}). 

Considering the above arguments on the necessity of the
Hubbard model generalization by taking into account orbital degeneration,
on the one hand, and correlated hopping, on the other hand, in the present
paper we propose a doubly orbitally degenerate narrow-band model  
with correlated hopping. The structure of this paper is the 
following. Section \ref{sec: Ham} is devoted to the model formulation and model 
Hamiltonian construction, the representation of the Hamiltonian of a 
doubly orbitally degenerate model with correlated hopping in the electron 
and Hubbard operators is given. In Section \ref{sec: MIT} metal-insulator transition in
the model at different integer values of the electron concentration is 
studied. The cases of the strong and weak Hund's coupling are 
considered. 
The absence of a natural small expansion parameter in the region near the
MIT requires to find the nonperturbative approaches; in such a situation
methods of mean-field type are useful. We shall use one of these methods,
a variant of the generalized Hartree-Fock approximation \cite{roth,zubar}
which has been proposed in Ref. \cite{1_6,2_19}. The approach gives the
exact band and atomic limits in the single-band Hubbard model and describes
MIT. The method reproduces (see Ref. \cite{1_62}) the exact results (criterion
of MIT (\ref{exact_cr}) and ground state energy) for the case of 
a half-filled non-degenerate band in a generalized Hubbard model with 
correlated hopping at $t_0=-X$.
By means of the generalized mean-field approximation 
the single-particle Green function and quasiparticle energy 
spectrum are calculated. With the help of the obtained energy spectrum we 
find energy gap width and criteria of
metal-insulator transition. Finally, Section \ref{sec: concl} is devoted to the 
conclusions and the comparison of our results with those of other authors. 

\section{Model Hamiltonian}
\label{sec: Ham}
\setcounter{equation}{0}

On the analogy of an orbitally non-degenerate model \cite{1_12,1_6} we start 
from the following generalization of the Hubbard model for an orbitally
degenerate band taking into account the matrix element of electron-electron
interaction which describes intersite hoppings of electrons (correlated
hopping):
\begin{eqnarray} \label{H1}
H=&&-\mu \sum_{i\gamma\sigma}a_{i\gamma\sigma}^{+}a_{i\gamma\sigma}+
{\sum\limits_{ij\gamma\sigma}}' a_{i\gamma\sigma}^{+}\left(
t_{ij}+\sum\limits_{k\gamma'}J(i\gamma k\gamma' j\gamma k\gamma')n_{k\gamma'}
\right) a_{j\gamma\sigma}
\nonumber\\
&&+U \sum_{i\gamma} n_{i\gamma\uparrow}n_{i\gamma\downarrow}
+U' \sum_{i\sigma}n_{i\alpha\sigma}n_{i\beta\bar{\sigma}}
+(U'-J)\sum_{i\sigma}n_{i\alpha\sigma}n_{i\beta\sigma},
\end{eqnarray}
where 
$\mu$ is the chemical potential, $a_{i\gamma\sigma}^{+}, a_{i\gamma\sigma}$ 
are the creation and destruction 
operators of an electron of spin $\sigma$ ($\sigma =\uparrow, \downarrow$;
${\bar \sigma}$ denotes spin projection which is opposite to $\sigma$) 
on $i$-site and in orbital $\gamma$ ($\gamma=\alpha ,\beta$ denotes two 
possible values of orbital states),  
$n_{i\gamma\sigma}=a_{i\gamma\sigma}^{+}a_{i\gamma\sigma}$
is the number operator of electrons of spin $\sigma$ and in orbital $\gamma$ 
on $i$-site, $n_{i\gamma}=n_{i\gamma\uparrow}+n_{i\gamma\downarrow}$;
$t_{ij}$ is the hopping integral of an electron
from $\gamma$-orbital of $j$-site to $\gamma$-orbital of $i$-site  
(we neglect the electron hoppings between $\alpha$- and $\beta$-orbitals).
In real systems the electron hoppings between 
different orbitals can exist, in addition the hopping integrals are anisotropic for 
$e_g$ orbitals. This may have an effect on the orbital and 
magnetic ordering~\cite{3,ish,shi}. We, however,
assume for simplicity $t^{\alpha\beta}_{ij}=t_{ij}\delta_{\alpha\beta}$.
This assumption considerably simplifies the analysis of properties of the 
model under consideration and allows to describe the physics of the 
metal-insulator transition.
\begin{eqnarray}
J(i\gamma k\gamma' j\gamma k\gamma')=\int\int\varphi^*_{\gamma}
({\bf r}-{\bf R}_i)\varphi_{\gamma}({\bf r}-{\bf R}_j)
{e^2\over |{\bf r}-{\bf r'}|}|\varphi_{\gamma'}({\bf r'}-{\bf R}_k)|^2
{\bf drdr'}
\end{eqnarray}
($\varphi_{\gamma}$ is the Wannier function), the prime at second 
sum in Eq.~(\ref{H1}) signifies that $i\not=j$,
\begin{eqnarray}
U=\int\int|\varphi_{\gamma}({\bf r}-{\bf R}_i)|^2
{e^2\over |{\bf r}-{\bf r'}|}|\varphi_{\gamma}({\bf r'}-{\bf R}_i)|^2
{\bf drdr'}
\end{eqnarray}
is the intra-atomic Coulomb repulsion of two electrons of the opposite spins
at the same orbital (we assume that it has the same value at $\alpha$- and 
$\beta$-orbitals),
\begin{eqnarray}
U'=\int\int|\varphi_{\alpha}({\bf r}-{\bf R}_i)|^2
{e^2\over |{\bf r}-{\bf r'}|}|\varphi_{\beta}({\bf r'}-{\bf R}_i)|^2
{\bf drdr'}
\end{eqnarray}
is the intra-atomic Coulomb repulsion of two electrons of the opposite spins
at the different orbitals,
\begin{eqnarray}
J=\int\int\varphi^*_{\alpha}({\bf r}-{\bf R}_i)
\varphi_{\beta}({\bf r}-{\bf R}_i)
{e^2\over |{\bf r}-{\bf r'}|}\varphi^*_{\beta}({\bf r'}-{\bf R}_i)
\varphi_{\alpha}({\bf r'}-{\bf R}_i) {\bf drdr'}
\end{eqnarray}
is the intra-atomic exchange interaction energy which stabilizes the Hund's
states 
forming the atomic magnetic moments.
The parameters $U,\ U',\ J$ are connected by the relation \cite{lacr}:
\begin{eqnarray}
U'=U-2J.
\end{eqnarray}

In Hamiltonian (\ref{H1}) we rewrite the sum 
${\sum\limits_{ijk\gamma\gamma'\sigma}}'
J(i\gamma k\gamma' j\gamma k\gamma')a_{i\gamma\sigma}^{+}n_{k\gamma'}
a_{j\gamma\sigma}$ in the form:
\begin{eqnarray} \label{c_h}
&&{\sum\limits_{ij\gamma\sigma}}'\left(J(i\gamma i\gamma j\gamma i\gamma)
a_{i\gamma\sigma}^{+}a_{j\gamma\sigma}n_{i\gamma\bar{\sigma}} + h.c.\right)+ 
{\sum\limits_{ij\gamma\sigma}}'\left(J(i\gamma i\bar{\gamma} j\gamma 
i\bar{\gamma}) 
a_{i\gamma\sigma}^{+}a_{j\gamma\sigma}n_{i\bar{\gamma}} + h.c.\right)
\nonumber\\
&&+{\sum\limits_{ij\gamma\gamma'\sigma}}'\sum_{\stackrel{k\ne i}{k\ne j}}
J(i\gamma k\gamma' j\gamma k\gamma')a_{i\gamma\sigma}^{+}a_{j\gamma\sigma}
n_{k\gamma'}
\end{eqnarray}
($\bar{\gamma}=\beta$ if $\gamma=\alpha$, and $\bar{\gamma}=\alpha$ when
$\gamma=\beta$).

The first and third terms of Eq. (\ref{c_h}) generalize correlated hopping
introduced for an orbitally non-degenerate model (e.g., see Ref \cite{1_6}).
The second term of expression (\ref{c_h}) describes correlated hopping of
electrons being the peculiarity of orbitally degenerate systems only (it is
absent in a single band case). Among this type of processes one can separate out
three essentially different (from the energy point of view) hoppings 
($X_i^{kl}$-representation allows to have done this easily). 
 
The first and second sums of Eq. (\ref{c_h}) describe the hoppings of 
electrons which are correlated by electron occupation of sites involved
in the hopping process. 
The third sum describes the hoppings of an electron between
$|i\gamma\sigma\rangle$- and $|j\gamma\sigma\rangle$-states which are 
dependent on the occupation number $n_k$ of other ($k\ne i,\ k\ne j$) sites. 
Let us take into account the influence of occupation of these sites
in the Hartree-Fock approximation:
\begin{eqnarray}
{\sum\limits_{ij\gamma\gamma'\sigma}}'\sum_{\stackrel{k\ne i}{k\ne j}}
J(i\gamma k\gamma' j\gamma k\gamma')a_{i\gamma\sigma}^{+}a_{j\gamma\sigma}
n_{k\gamma'}\simeq n{\sum\limits_{ij\gamma\sigma}}'T_1(ij)
a_{i\gamma\sigma}^{+}a_{j\gamma\sigma},
\end{eqnarray}
where $n=\langle n_{i\alpha}+n_{i\beta}\rangle$ is the average number of 
electrons per site,
\begin{eqnarray}
T_1(ij)=\sum_{\stackrel{k\ne i}{k\ne j}}J(i\gamma k\gamma' j\gamma k\gamma')
\end{eqnarray}
(we suppose $J(i\gamma k\alpha j\gamma k\alpha)=
J(i\gamma k\beta j\gamma k\beta)$ and $T_1(ij)$ have the same value for both 
$\alpha$- and $\beta$-orbitals). Assuming that $\alpha$- and $\beta$-states 
are equivalent, denote:
\begin{eqnarray} 
&&J(i\gamma i\bar{\gamma} j\gamma i\bar{\gamma})=t'_{\alpha\alpha}(ij)=
t'_{\beta\beta}(ij)=t'_{ij},
\\
&&J(i\gamma i\gamma j\gamma i\gamma)=t''_{\alpha\alpha}(ij)=
t''_{\beta\beta}(ij)=t''_{ij}.
\end{eqnarray}

So we can rewrite Hamiltonian (\ref{H1}) in the following form:
\begin{eqnarray} \label{H2}
H=&&-\mu \sum_{i\gamma\sigma}a_{i\gamma\sigma}^{+}a_{i\gamma\sigma}+
{\sum\limits_{ij\gamma\sigma}}'t_{ij}(n) a_{i\gamma\sigma}^{+}
a_{j\gamma\sigma}+{\sum\limits_{ij\gamma\sigma}}'(t'_{ij}
a_{i\gamma\sigma}^{+}a_{j\gamma\sigma}n_{i\bar{\gamma}} + h.c.)
\nonumber\\
&&+{\sum\limits_{ij\gamma\sigma}}'(t''_{ij}
a_{i\gamma\sigma}^{+}a_{j\gamma\sigma}n_{i\gamma\bar{\sigma}} + h.c.)
+U \sum_{i\gamma} n_{i\gamma\uparrow}n_{i\gamma\downarrow}
+U' \sum_{i\sigma}n_{i\alpha\sigma}n_{i\beta\bar{\sigma}}
\\
&&+(U'-J)\sum_{i\sigma}n_{i\alpha\sigma}n_{i\beta\sigma},
\nonumber
\end{eqnarray}
with the effective hopping integral $t_{ij}(n)=t_{ij}+nT_1(ij)$ being the 
concentration-dependent in consequence of taking into account correlated
hopping $T_1(ij)$.

The distinction of Hamiltonian (\ref{H2}) from models of narrow-band 
materials with orbital degeneracy is taking into account the matrix element
$J(i\gamma k\gamma' j\gamma k\gamma')$ caused by electron-electron 
interaction.
This leads to the electron-hole asymmetry (which is analogous one to the
case of an non-degenerate band \cite{1_25,1_6}) and the dependence 
of hopping integral on the average number of electrons per site. Thus the 
narrow-band model which are described by Hamiltonian (\ref{H2}) shows
much better physics than the Hubbard model with doubly
orbital degeneration.

In the model described by Hamiltonian~(\ref{H2}) each site of the 
lattice can be in one of 16 states (see Fig.~\ref{sites}). 

Let us pass to the configurational representation of Hamiltonian (\ref{H2}). 
Use the representation of the operators of creation and destruction 
of electron through $X_i^{kl}$-operators of transition of site $i$
from the state $l$ to the state $k$:
\begin{eqnarray}
a_{i\alpha\uparrow}^{+}=X_i^{\alpha\uparrow ,0}
			+X_i^{\uparrow\uparrow,\beta\uparrow}
			+X_i^{\uparrow\downarrow,\beta\downarrow}
			+X_i^{\alpha2,\alpha\downarrow}
			+X_i^{\alpha2\downarrow,\downarrow\downarrow}
			+X_i^{\alpha2\uparrow,\downarrow\uparrow}
			+X_i^{\beta2\uparrow,\beta2}
			+X_i^{4,\beta2\downarrow}, \ \
\nonumber\\ 
a_{i\alpha\downarrow}^{+}=X_i^{\alpha\downarrow ,0}
			+X_i^{\downarrow\downarrow,\beta\downarrow}
 			+X_i^{\downarrow\uparrow,\beta\uparrow}
			-X_i^{\alpha2,\alpha\uparrow}
			-X_i^{\alpha2\downarrow,\uparrow\downarrow}
			-X_i^{\alpha2\uparrow,\uparrow\uparrow}
			+X_i^{\beta2\downarrow,\beta2}
			-X_i^{4,\beta2\uparrow}, \ \
\\ 
a_{i\beta\uparrow}^{+}=X_i^{\beta\uparrow ,0}
			-X_i^{\uparrow\uparrow,\alpha\uparrow}
			-X_i^{\downarrow\uparrow,\alpha\downarrow}
			+X_i^{\beta2,\beta\downarrow}
			-X_i^{\beta2\downarrow,\downarrow\downarrow}
			-X_i^{\beta2\uparrow,\uparrow\downarrow}
			+X_i^{\alpha2\uparrow,\alpha2}
			+X_i^{4,\alpha2\downarrow}, \ \
\nonumber \\
a_{i\beta\downarrow}^{+}=X_i^{\beta\downarrow ,0}
			-X_i^{\downarrow\downarrow,\alpha\downarrow}
			-X_i^{\uparrow\downarrow,\alpha\uparrow}
			-X_i^{\beta2,\beta\uparrow}
			+X_i^{\alpha2\downarrow,\alpha2}
			+X_i^{\beta2\uparrow,\uparrow\uparrow}
			+X_i^{\beta2\downarrow,\downarrow\uparrow}
			-X_i^{4,\alpha2\uparrow}. \ \
\nonumber
\end{eqnarray}
Such choice of representation with taking into account the 
commutational rules
\begin{eqnarray}
&&\{X_i^{p,l};X_j^{k,t}\}=\delta_{ij}(\delta_{lk}X_i^{p,t}
+\delta_{tp}X_i^{k,l}),
\nonumber\\
&&[X_i^{p,l};X_j^{k,t}]=\delta_{ij}(\delta_{lk}X_i^{p,t}
-\delta_{tp}X_i^{k,l}),
\end{eqnarray}
where the anticommutator ($\{A; B\}$) has to be taken only if both operators are
fermionic, i.e., change the particle number by one (e.g. 
$X_{i}^{\gamma\sigma,0}$ or $X_i^{\gamma 2\sigma,4}$),
and the constraint
\begin{eqnarray}
\sum_{p}X_i^{p}=1,
\end{eqnarray}
where
$X_i^p=X_i^{p,l}X_i^{l,p}$  
is the operator of number of $|p\rangle$ states on the site $i$
ensures the fulfillment of the  anticommutation relations
for $a$-operators.

The model Hamiltonian in the configurational representation has the form:
\begin{eqnarray} \label{H_x}
H=&&-\mu \left(\sum_{i\sigma}X_i^{\gamma\sigma}
+2 \sum_{i\sigma}(X_i^{\sigma\sigma}+X_i^{\sigma\bar{\sigma}})
+2 \sum_{i\gamma}X_i^{\gamma2}
+3 \sum_{i\gamma\sigma}X_i^{\gamma2\sigma}
+4 \sum_{i}X_i^{4} \right) 
\nonumber \\
&&+(U'-J)\sum_{i\sigma}X_i^{\sigma\sigma}
+U'\sum_{i\sigma}X_i^{\sigma\bar{\sigma}}
+U\sum_{i\gamma}X_i^{\gamma2} \\
&&+(U+2U'-J)\sum_{i\gamma\sigma}X_i^{\gamma2\sigma}
+2(U+2U'-J)\sum_{i}X_i^{4}  
+H_{t},
\nonumber
\end{eqnarray} 
where the kinetic part of the Hamiltonian is
\begin{eqnarray}
H_{t}=\sum_{n,m}H_{nm}  
\end{eqnarray}
with $n,m = \{0$-$\gamma\sigma,\ \gamma\sigma$-$\sigma\sigma,\ 
\gamma\sigma$-$\sigma{\bar\sigma},\ \gamma\sigma$-$\gamma 2,\ 
\sigma\sigma$-$\gamma 2\sigma,\ \sigma{\bar\sigma}$-$\gamma 2\sigma,\ 
\gamma 2$-$\gamma 2\sigma,\ \gamma 2\sigma$-$4\}$. 

The Hamiltonians $H_{nm}$ contain  
the processes which form the energy subbands (analogues of the
Hubbard subbands) and the processes of the hybridization of these subbands.
In Fig.~\ref{levels} the transitions between the states of the sites which form the
corresponding subbands are shown.
 
 The different hopping integrals $t_{ij}^{nm}$
correspond to the transitions within the different subbands $H_{nn}$ or 
between the different subbands $H_{nm} \ (n\ne m)$:
\begin{eqnarray}
t_{ij}^{nm}=t_{ij}+(\tau^{n}+\tau^{m}),
\end{eqnarray}
where
\begin{eqnarray}
&&\tau^{0-\gamma\sigma}=0,       \hspace{24mm}  
\tau^{\sigma\sigma-\gamma 2\sigma}=t'_{ij}+t''_{ij}, 
\nonumber \\
&&\tau^{\gamma\sigma-\sigma\sigma}=t'_{ij},  \hspace{20mm}  
\tau^{\sigma{\bar\sigma}-\gamma 2\sigma}=t'_{ij}+t''_{ij},
\nonumber \\
&&\tau^{\gamma\sigma-\sigma{\bar\sigma}}=t'_{ij}, \hspace{20mm}  
\tau^{\gamma 2-\gamma 2\sigma}=2t'_{ij}, 
\\
&&\tau^{\gamma\sigma-\gamma 2}=t''_{ij},\hspace{20mm}  
\tau^{\gamma 2\sigma-4}=2t'_{ij}+t''_{ij}.
\nonumber 
\end{eqnarray}

The mutual placement and the overlapping of the noted subbands depend on the relations
between the values of intra-atomic Coulomb repulsion parameters $U, U'$, 
intra-atomic exchange interaction parameter $J$ and the widths of subbands.

\section{Metal-insulator transition}
\label{sec: MIT}
\setcounter{equation}{0}

At integer values of electron concentration $n=1,\ 2,\ 3$ in the model
described by the Hamiltonian~(\ref{H_x})  MIT can occur. 
The possible metal-insulator transitions at
some integer values of the mean electron number per site will be 
considered below.

\subsection{Metal-insulator transition at electron concentration $n=1$}
\subsubsection{The strong Hund's coupling case }

Let us consider the case of 
the strong intra-atomic Coulomb interaction $U' \gg t_{ij}$ and the
strong Hund's  coupling $U' \gg U'-J$ (values $U'$ and $J$ are of the 
same order). These conditions allow us to neglect the states of site when
there are more than two electrons on the site and the ``non-Hund's''
doubly occupied states 
$|\!\uparrow\!\downarrow \rangle$, $|\!\downarrow\!\uparrow\rangle $, 
$|\alpha2\rangle$, $|\beta2\rangle$ 
(the analogous conditions are used for an investigation of magnetic 
properties of the Hubbard model with twofold orbital 
degeneration in Ref.~\cite{lacr,kubo,nak}).
Thus lattice sites can be in one of seven possible states:
a hole (a non-occupied by electron site) $|0\rangle$; 
a single occupied by electron site $|\alpha\!\!\uparrow\rangle$, 
$|\alpha\!\!\downarrow\rangle$, $|\beta\!\!\uparrow\rangle$, 
$|\beta\!\!\downarrow\rangle$;
the Hund's doublon or site with two electrons on the different orbitals
with the same spins $|\uparrow\!\uparrow\rangle$, 
$|\downarrow\!\downarrow\rangle$.

Let us pass to the configurational representation of the Hamiltonian.
Electron operators in terms of the transition operators $X_i^{k,l}$  
of site $i$ from 
the state $|l\rangle$ to the state $|k\rangle$ are expressed by the 
formulae:
\begin{eqnarray}
a_{i\alpha\uparrow}^{+}=X_i^{\alpha\uparrow ,0}+X_i^{\uparrow\uparrow
,\beta\uparrow}, \ \
a_{i\alpha\uparrow}=X_i^{0,\alpha\uparrow}+X_i^{\beta\uparrow ,\uparrow
\uparrow},
\nonumber\\
a_{i\alpha\downarrow}^{+}=X_i^{\alpha\downarrow ,0}+X_i^{\downarrow
\downarrow ,\beta\downarrow}, \ \
a_{i\alpha\downarrow}=X_i^{0, \alpha\downarrow}+X_i^{\beta\downarrow ,
\downarrow\downarrow},
\\
a_{i\beta\uparrow}^{+}=X_i^{\beta\uparrow ,0}-X_i^{\uparrow\uparrow ,
\alpha\uparrow}, \ \
a_{i\beta\uparrow}=X_i^{0,\beta\uparrow}-X_i^{\alpha\uparrow ,\uparrow
\uparrow},
\nonumber\\
a_{i\beta\downarrow}^{+}=X_i^{\beta\downarrow ,0}-X_i^{\downarrow
\downarrow ,\alpha\downarrow}, \ \
a_{i\beta\downarrow}=X_i^{0,\beta\downarrow}-X_i^{\alpha\downarrow
 ,\downarrow\downarrow},
\nonumber
\end{eqnarray}
(in these formulae we have took into account only seven possible states).
It should be emphasized that one have to consider all the possible site states
in order to get the correct anticommutation rules for electron operators.

Let us calculate quasiparticle energy spectrum of the model neglecting
the correlated hopping $t'_{ij}=t''_{ij}=0,\ t_{ij}(n)=t_{ij}$.
We write the Hamiltonian in the form:
\begin{eqnarray}
H=&&-\mu \sum_{i\sigma}\left(X_i^{\alpha\sigma}+X_i^{\beta\sigma}+2X_i^
{\sigma\sigma}\right)+
{\sum \limits_{ij\gamma\sigma}}'t_{ij}\left(X_i^
{\gamma\sigma ,0} X_j^{0,\gamma\sigma} +
X_i^{\sigma\sigma ,\gamma\sigma}X_j^{\gamma\sigma ,\sigma
\sigma}\right)+
\nonumber\\
&&+(U'-J)\sum_{i\sigma}X_i^{\sigma\sigma}+H_{\alpha\beta},
\end{eqnarray}
where 
\begin{eqnarray}
H_{\alpha\beta}=&&\sum \limits_{ij}t_{ij}\Bigl(-X_i^{\uparrow\uparrow
,\alpha\uparrow}X_j^{0, \beta\uparrow} +X_i^{\uparrow\uparrow
,\beta\uparrow} X_j^{0, \alpha\uparrow}
+ X_i^{\downarrow \downarrow
,\beta \downarrow}X_j^{0, \alpha\downarrow} -
\nonumber\\
&&X_i^{\downarrow\downarrow
,\alpha\downarrow}X_j^{0, \beta\downarrow}+h.c.\Bigr).
\end{eqnarray}

For calculation of quasiparticle energy spectrum and polar state 
concentration $c\equiv \langle   X_i^0\rangle
$ -- holes, $d_{\uparrow}\equiv \langle   X_i^
{\uparrow\uparrow}\rangle$, $d_{\downarrow}\equiv \langle   X_i^
{\downarrow\downarrow}\rangle$  being the Hund's doublons
we use the method of the retarded Green functions.
For the Green function
\begin{eqnarray} 
G_{pp'}^{(1)}(E)=\langle\langle X_p^{\alpha\uparrow, 0\!}|\!X_{p'}^{0, \alpha\uparrow}
\rangle\rangle
\end{eqnarray}
we have the equation:
\begin{eqnarray} \label{gr1}
(E-\mu)\langle\langle X_p^{\alpha\uparrow,0\!}|\!X_{p'}^{0, \alpha\uparrow }
\rangle\rangle =&&{\delta_{pp'}\over 2\pi}\langle X_p^0+X_p^{\alpha\uparrow}\rangle
\nonumber\\
&&+\langle\langle\left[X_p^{\alpha\uparrow,0},H_0 \right] +
\left[X_p^{\alpha\uparrow,0},H_{\alpha\beta} \right]|\!X_{p'}^{0, 
\alpha\uparrow }\rangle\rangle,
\end{eqnarray}
where
\begin{eqnarray}
H_0={\sum \limits_{ij\gamma\sigma}}'t_{ij}\left(X_i^{\gamma\sigma ,0} X_j^{0,
\gamma\sigma} + X_i^{\sigma\sigma ,\gamma\sigma}X_j^{\gamma\sigma ,\sigma
\sigma}\right).
\end{eqnarray}

To break off the sequence of equations for Green functions 
we apply a variant~\cite{2_19} of the generalized
mean field approximation~\cite{roth,zubar}:
\begin{eqnarray} \label{apr}
&&\left[X_p^{\alpha\uparrow, 0},H_0\right]={\sum
\limits_j}'\epsilon (pj) X_j^{\alpha\uparrow,0},
\nonumber\\
&&\left[X_p^{\alpha\uparrow,0},H_{\alpha\beta} \right]={\sum
\limits_j}'\epsilon_1 (pj) X_j^{\uparrow\uparrow,\beta\uparrow},
\end{eqnarray}
where $\epsilon (pj),\ \epsilon_1 (pj)$ are the non-operator 
expressions, the method of its calculation will be given below.
The representation~(\ref{apr}) of commutators in Eq.~(\ref{gr1}) 
is defined by the operator structure of $H_0$ and $H_{\alpha\beta}$. 

Using the approximation~(\ref{apr}) Eq.~(\ref{gr1}) for the Green
function $G^{(1)}_{pp'}(E)$ can be written as
\begin{eqnarray} \label{eq1}
(E-\mu)\langle\langle X_p^{\alpha\uparrow,0\!}|\!X_{p'}^{0, \alpha\uparrow }
\rangle\rangle =&&{\delta_{pp'}\over 2\pi}\langle X_p^0+X_p^{\alpha\uparrow}\rangle
+{\sum\limits_j}'\epsilon (pj) \langle\langle 
X_j^{\alpha\uparrow,0}|\!X_{p'}^{0,\alpha\uparrow}\rangle\rangle
\nonumber\\
&&+{\sum\limits_j}'\epsilon_1 (pj) 
\langle\langle X_j^{\uparrow\uparrow,\beta\uparrow}|\!X_{p'}^{0, \alpha\uparrow}
\rangle\rangle.
\end{eqnarray}

For the Green function
\begin{eqnarray}
G_{pp'}^{(2)}(E)=\langle\langle X_p^{\uparrow\uparrow,\beta\uparrow}\!
|\!X_{p'}^{0,\alpha\uparrow } \rangle\rangle
\end{eqnarray}
we write the analogous equation:
\begin{eqnarray} \label{eq2}
(E-\mu+U'-J)\langle\langle X_p^{\uparrow\uparrow, \beta\uparrow}\!|\!X_{p'}^{0,\alpha
\uparrow }\rangle\rangle =&&{\sum\limits_j}'\tilde{\epsilon}(pj)\langle\langle X_j^{
\uparrow\uparrow, \beta\uparrow}\!|\!X_{p'}^{0,\alpha\uparrow }\rangle\rangle
\nonumber\\
&&+{\sum\limits_j}'\epsilon_2(pj)\langle\langle X_j^{\alpha\uparrow, 0}\!|\!X_{p'}
^{0,\alpha\uparrow}\rangle\rangle,
\end{eqnarray}
where the non-operator expressions $\tilde{\epsilon},\ \epsilon_2$ 
are defined by the formulae:
\begin{eqnarray} \label{apr2}
&&\left[X_p^{\uparrow\uparrow, \beta\uparrow},H_0\right]={\sum
\limits_j}'\tilde{\epsilon} (pj) X_j^{\uparrow\uparrow, \beta\uparrow},
\nonumber\\
&&\left[X_p^{\uparrow\uparrow,\beta\uparrow},H_{\alpha\beta} \right]={\sum
\limits_j}'\epsilon_2 (pj) X_j^{\alpha\uparrow, 0}.
\end{eqnarray}
Thus, we obtain the closed system of equations for functions
$G_{pp'}^{(1)}(E)$ and $G_{pp'}^{(2)}(E)$.

To calculate  $\epsilon (pj),\ \epsilon_1 (pj),\ \tilde{\epsilon}(pj),\
\epsilon_2 (pj)$  we use the procedure proposed in paper~\cite{2_19}.  
The values of 
$\epsilon (pj),\ \epsilon_1 (pj),\ \tilde{\epsilon}(pj),\
\epsilon_2 (pj)$ 
we find by anticommutation of Eqs.(\ref{apr}), (\ref{apr2}) with the 
operators
$X_{p'}^{0,\alpha \! \uparrow\!}$ and $X_{p'}^{\beta\uparrow, 
\uparrow\uparrow}$ respectively:
\begin{eqnarray}
(X_p^{0}+X_{p}^{\alpha\uparrow})\epsilon(pp')&=&
\left\{X_{p'}^{0,\alpha \! \uparrow\!};
\left[X_p^{\alpha\uparrow, 0},H_0\right]\right\},
\nonumber\\
(X_{p}^{\beta \! \uparrow}
+X_{p}^{\uparrow \! \uparrow})
\epsilon_1(pp')&=&\left\{X_{p'}^{\beta\uparrow, \uparrow\uparrow};
\left[X_p^{\alpha\uparrow,0},H_{\alpha\beta} \right]\right\},
\nonumber\\
(X_{p}^{\beta \! \uparrow}
+X_{p}^{\uparrow \! \uparrow})
\tilde{\epsilon}(pp')&=&\left\{X_{p'}^{\beta\uparrow, 
\uparrow\uparrow};
\left[X_p^{\uparrow\uparrow, \beta\uparrow},H_0\right]\right\},
\\
(X_p^{0}+X_{p}^{\alpha\uparrow})
\epsilon_2 (pp')&=&\left\{X_{p'}^{0\!, \alpha\uparrow};
\left[X_p^{\uparrow\uparrow,\beta\uparrow},H_{\alpha\beta} \right]\right\}.
\nonumber
\end {eqnarray}

We rewrite $X_i^{kl}$-operator in the form 
$X_i^{k,l}=\alpha_{ik}^{+}\alpha_{il}$,
where $\alpha_{ik}^{+},\ \alpha_{il}$ are the operators of creation and 
destruction for $|k\rangle$- and $|l\rangle$-states
on $i$-site respectively (with the constraint
${\sum \limits_{k}}\alpha_{ik}^{+}\alpha_{ik}=1$); 
thus $X_i^0=\alpha^{+}_{i0}\alpha_{i0},\ 
X_i^{\sigma\sigma}=\alpha^{+}_{i\sigma\sigma}\alpha_{i\sigma\sigma},\ 
X_i^{\gamma\sigma}=\alpha^{+}_{i\gamma\sigma}\alpha_{i\gamma\sigma}$.
Let us substitute $\alpha$-operators by $c$-numbers 
(here there is a partial equivalence with the slave boson method~\cite{1_53}).
At $n=1$ in a paramagnetic state 
\begin{eqnarray}
&&d\equiv d_{\uparrow}=
d_{\downarrow}=c/2, 
\quad
\langle X_i^{\alpha\downarrow}\rangle=0.25-d,
\nonumber\\
&&\alpha^{+}_{i0}=\alpha_{i0}=c^{1/2};\quad \alpha^{+}_{i\sigma\sigma}
=\alpha_{i\sigma\sigma}=d^{1/2};\quad 
\alpha^{+}_{i\gamma\sigma}=\alpha_{i\gamma\sigma}=(0.25-d)^{1/2}.
\end{eqnarray}

 The proposed approximation is based on the following physical idea. 
When the lower and upper Hubbard subbands overlap weakly (the case of 
a paramagnetic Mott-Hubbard semimetal) 
$|0\rangle$- and $|\sigma\sigma\rangle$-
states almost do not influence on 
$|\gamma\sigma\rangle$-states then $|\gamma\sigma\rangle$-subsystem 
can be considered as a quasiclassical subsystem (an analogue
of the thermodynamic reservoir). In this case one can substitute 
$\alpha$-operators through $c$-numbers. We extend this ideology 
to determination of 
$\epsilon (pj),\ \epsilon_1 (pj),\ \tilde{\epsilon}(pj),\ \epsilon_2 (pj)$.
Thus we substitute the creation and destruction 
operators of $|0\rangle$-, $|\sigma\sigma\rangle$- and 
$|\gamma\sigma\rangle$-states through  respective quasiclassical expressions. 
Actually the proposed procedure is equivalent to a separation 
of the charge and spin degrees of freedom.

Then we obtain the formulae in ${\bf k}$-representation:
\begin{eqnarray}
\epsilon({\bf k})={-0.5c^2+0.5c-0.25 \over {0.25+0.5c}} t_{\bf k}, \qquad
\epsilon_1({\bf k})=(1.5c+c^2)t_{\bf k},
\nonumber\\
\tilde{\epsilon}({\bf k})=(c^2+c-0.5) t_{\bf k}, \qquad
\epsilon_2({\bf k})={0.25c^2+0.375c \over {0.25+0.5c}}t_{\bf k}.
\end{eqnarray}

From system of Eqs.~(\ref{eq1}), (\ref{eq2}) we have
\begin{eqnarray}\label{G1}
&&G_{\bf k}^{(1)}(E)={0.5c+0.25 \over {2\pi}}
{{E-\mu +U'-J-\tilde{\epsilon}(\bf k)}
\over {(E+E_1({\bf k}))(E+E_2({\bf k}))}},
\\
&&G_{\bf k}^{(2)}(E)={0.5c+0.25 \over 2\pi}
{\epsilon_2({\bf k})
\over (E+E_1({\bf k}))(E+E_2({\bf k}))},
\end{eqnarray}
where the expressions
\begin{eqnarray} \label{E1}
E_{1,2}({\bf k})=&&{-\mu }-{\epsilon({\bf k})-
\tilde{\epsilon}({\bf k})+(U'-J)\over 2} \mp
{1\over 2} \biggl\{ [
\epsilon({\bf k})+\tilde{\epsilon}({\bf k})-(U'-J)]^2
\nonumber\\
&&+4 \epsilon({\bf k})(U'-J)+4[\epsilon_1({\bf k})
\epsilon_2({\bf k})-
\epsilon({\bf k}) \tilde{\epsilon}({\bf k})] \biggr\}^{1/2}
\end{eqnarray}
are the quasiparticle energy spectrum.
$E_2({\bf k})$ describes the electron spectrum in  
the 0-$\alpha\sigma$-subband (an analogue of the 
lower Hubbard subband), $E_1({\bf k})$ describes the electron spectrum in the 
$\alpha\sigma$-$\sigma\sigma$-subband (an analogue of
the upper Hubbard subband). 
The computation of the quasiparticle energy spectrum in the 
0-$\beta\sigma$- and $\beta\sigma$-$\sigma\sigma$-subbands gives
the same Eqs.~(\ref{E1}).  

The energy gap width (difference of energies between bottom of the upper 
and top of the lower bands) is
\begin{eqnarray}\label{dE}
\Delta E=E_2(-w)-E_1(w).
\end{eqnarray}

The peculiarity of the obtained energy spectrum and energy gap width
is their dependence on the hole concentration $c$. With rise of
temperature or the parameter $(U-J)/(2w)$ the holes concentration
decrease smoothly.

 The energy gap width $\Delta E$ as a function of the parameters
$(U'-J)/(2w)$ and $(kT)/(2w)$ is presented in Fig.~\ref{gap_w} 
and Fig.~\ref{gap_t} respectively.
With a change of the parameter $(U'-J)/(2w)$ system undergoes the transition 
from an insulating to a metallic state (negative values of the energy gap
width correspond to the overlapping of the Hubbard subbands).
In the model under consideration  at $T=0$~K insulator-metal transition 
at $n=1$ occurs when $(U'-J)/(2w)=0.75$ (Fig.~\ref{gap_w}, the lower curve)
(in the single-band Hubbard model the respective parameter is 
$U/(2w)=1$~\cite{2_19}).

The transition from a metallic to an insulating state with increase 
of temperature at given value of the parameter $(U'-J)/(2w)$ is also possible
(Fig.~\ref{gap_t}). It can be explained by the fact that energy gap width $\Delta E$
(\ref{dE}) increases at increasing temperature $T$ which is caused
by the rise of polar states concentration at constant 
$w$, $(U'-J)$. 

\subsubsection{The limit of the weak Hund's coupling}

Let us consider the MIT at electron concentration $n=1$.
If the exchange interaction is small comparatively to the Coulomb
interaction $J \ll U$ then we can take $J$ into account in
the mean-field approximation (see, e.g., Ref.~\cite{5}). The corresponding 
terms of the Hamiltonian (\ref{H2}) can be written as:
\begin{eqnarray} \label{HF}
\nonumber
&&-2J \sum_{i\sigma}n_{i\alpha\sigma}n_{i\beta\bar{\sigma}}
-3J\sum_{i\sigma}n_{i\alpha\sigma}n_{i\beta\sigma}= 
-2J \sum_{i\sigma}(\langle n_{i\alpha\sigma}\rangle n_{i\beta\bar{\sigma}}\\
&&+n_{i\alpha\sigma}\langle n_{i\beta\bar{\sigma}}\rangle)
-3J\sum_{i\sigma}(\langle n_{i\alpha\sigma}\rangle n_{i\beta\sigma}+
n_{i\alpha\sigma}\langle n_{i\beta\sigma}\rangle)=\\
&&-5J\langle n_{i\gamma\sigma}\rangle \sum_{i\gamma\sigma}n_{i\gamma\sigma}.
\nonumber
\end{eqnarray}
 The Hamiltonian (\ref{H2}) takes the form: 
\begin{eqnarray} \label{H3}
H=&&-\tilde{\mu} \sum_{i\gamma\sigma}a_{i\gamma\sigma}^{+}a_{i\gamma\sigma}+
{\sum\limits_{ij\gamma\sigma}}'t_{ij}(n) a_{i\gamma\sigma}^{+}
a_{j\gamma\sigma}
\nonumber\\
&&+{\sum\limits_{ij\gamma\sigma}}'(t'_{ij}
a_{i\gamma\sigma}^{+}a_{j\gamma\sigma}n_{i\bar{\gamma}} + h.c.)
+{\sum\limits_{ij\gamma\sigma}}'(t''_{ij}
a_{i\gamma\sigma}^{+}a_{j\gamma\sigma}n_{i\gamma\bar{\sigma}} + h.c.)
\\
&&+U \sum_{i\sigma\gamma} (n_{i\gamma\sigma}n_{i\gamma\bar{\sigma}}
+n_{i\gamma\sigma}n_{i\bar{\gamma}\bar{\sigma}}
+n_{i\gamma\sigma}n_{i\bar{\gamma}\sigma}),
\nonumber
\end{eqnarray}
where $\tilde{\mu}=\mu+5J\langle n_{i\gamma\sigma}\rangle$.

Considering MIT at the electron concentration $n$ we can 
take into account in the Hamiltonian 
only the states of site with $n-1,\ n,\ n+1$ 
electrons (the analogous simplification has been used in Refs.~\cite{12,7}).
In the vicinity of the transition point at the electron concentration 
$n=1$ the concentrations of sites occupied by three and four electrons
are small. Neglecting the small amounts of these sites we can write 
the electron operators in the form:
\begin{eqnarray}
a_{i\alpha\uparrow}^{+}=X_i^{\alpha\uparrow ,0}
			+X_i^{\uparrow\uparrow,\beta\uparrow}
			+X_i^{\uparrow\downarrow,\beta\downarrow}
			+X_i^{\alpha2,\alpha\downarrow},
\nonumber\\ 
a_{i\alpha\downarrow}^{+}=X_i^{\alpha\downarrow ,0}
			+X_i^{\downarrow\downarrow,\beta\downarrow}
			+X_i^{\downarrow\uparrow,\beta\uparrow}
			-X_i^{\alpha2,\alpha\uparrow},\\ 
a_{i\beta\uparrow}^{+}=X_i^{\beta\uparrow ,0}
			-X_i^{\uparrow\uparrow,\alpha\uparrow}
			-X_i^{\downarrow\uparrow,\alpha\downarrow}
			+X_i^{\beta2,\beta\downarrow},
\nonumber \\
a_{i\beta\downarrow}^{+}=X_i^{\beta\downarrow ,0}
			-X_i^{\downarrow\downarrow,\alpha\downarrow}
			-X_i^{\uparrow\downarrow,\alpha\uparrow}
			-X_i^{\beta2,\beta\uparrow}.
\nonumber
\end{eqnarray}
Hamiltonian (\ref{H3}) in the configurational representation takes the form:
\begin{eqnarray}
H&=&-\tilde{\mu} \left(\sum_{i\gamma\sigma}X_i^{\gamma\sigma}
+2 \sum_{i\sigma}(X_i^{\sigma\sigma}+X_i^{\sigma\bar{\sigma}})
+2 \sum_{i\gamma}X_i^{\gamma2}\right)
\nonumber \\
&+&U\left(\sum_{i\sigma}X_i^{\sigma\sigma}
+\sum_{i\sigma}X_i^{\sigma\bar{\sigma}}
+\sum_{i\gamma}X_i^{\gamma2}\right)+H_1+H_2,
\end{eqnarray} 
where the kinetic part of the Hamiltonian is 
\begin{eqnarray}
H_1=&&\sum_{i,j}\Biggl(t_{ij}\sum_{\gamma \sigma}X_i^{\gamma \sigma,0}
X_j^{0,\gamma \sigma}
+(t_{ij}+2t''_{ij})\sum_{\gamma \sigma}X_i^{\gamma2,\gamma\sigma}
X_j^{\gamma\sigma,\gamma2}
\nonumber\\
&&+(t_{ij}+2t'_{ij})\biggl(\sum_{\gamma \sigma}
X_i^{\sigma \sigma,\gamma \sigma}
X_j^{\gamma \sigma,\sigma \sigma}
+\sum_{\sigma}\Bigl(X_i^{\sigma {\bar\sigma},\alpha \sigma}
X_j^{\alpha \sigma,\sigma {\bar\sigma}}
\nonumber\\
&&+X_i^{\sigma {\bar\sigma},\beta {\bar \sigma}}
X_j^{\beta {\bar \sigma},\sigma {\bar\sigma}}
+(X_i^{\sigma\sigma,\alpha\sigma}
X_j^{\alpha{\bar\sigma},{\bar\sigma} \sigma}
+X_i^{\sigma\sigma,\beta\sigma}
X_j^{\beta{\bar\sigma},\sigma{\bar\sigma}}+h.c.)\Bigr)\biggr)
\\
&&+(t_{ij}+t'_{ij}+t''_{ij})
\Bigl( \sum_{\gamma \sigma}\eta_{\sigma}\eta_{{\bar\gamma}}
X_i^{\sigma\sigma,\gamma\sigma}
X_j^{{\bar\gamma}{\bar\sigma},{\bar\gamma}2}
\nonumber\\
&&+\sum_{\sigma}\eta_{\sigma}
(X_i^{\sigma{\bar\sigma},\beta{\bar\sigma}}
X_j^{\alpha{\bar\sigma},\alpha2}+
X_i^{\sigma{\bar\sigma},\alpha\sigma}
X_j^{\beta\sigma,\beta2})+h.c.\Bigr) \Biggr),
\nonumber 
\end{eqnarray}
\begin{eqnarray}
H_2=\sum_{i,j}\Biggl(
(t_{ij}+t'_{ij})
\Bigl(\sum_{\gamma \sigma}(\eta_{\gamma}
X_i^{\gamma\sigma,0}
X_j^{{\bar\gamma}\sigma,\sigma\sigma}+ h.c.)
\nonumber\\
\sum_{\sigma}
(X_i^{\alpha \sigma,0}X_j^{\beta {\bar\sigma},\sigma {\bar\sigma}}-
X_i^{\beta \sigma,0}X_j^{\alpha{\bar\sigma},{\bar\sigma}\sigma}+h.c.)\Bigr)
\\
+(t_{ij}+t''_{ij})\sum_{\gamma \sigma}(\eta_{\sigma}
X_i^{\gamma \sigma,0}
X_j^{\gamma {\bar\sigma},\gamma2} +h.c.)\Biggr),
\nonumber     
\end{eqnarray}
where $\eta_{\uparrow}=\eta_{\alpha}=1,\ \eta_{\downarrow}=\eta_{\beta}=-1$.

The processes that form energy subbands are included in the 
Hamiltonian $H_1$, the processes  of 
hybridization of these subbands are included in the Hamiltonian $H_2$.

 The single-particle Green function can be written as:
\begin{eqnarray}
\langle\langle a_{p\alpha\uparrow}|a^{+}_{p'\alpha\uparrow}\rangle\rangle=
\langle\langle X_p^{0, \alpha \! \uparrow\!}|\!X_{p'}^{\alpha \! \uparrow,0}\rangle\rangle
+\langle\langle X_p^{0, \alpha \! \uparrow\!}|\!Y^+_{p'}\rangle\rangle \\
+\langle\langle Y_p|\!X_{p'}^{\alpha \! \uparrow,0}\rangle\rangle
+\langle\langle Y_p|\!Y^+_{p'}\rangle\rangle.
\nonumber
\end{eqnarray}
Here we have introduced the following notation:  
$Y_p=X_{p}^{\alpha \! \downarrow,\alpha2}
+X_{p}^{\beta \! \uparrow,\uparrow \! \uparrow}
+X_{p}^{\beta \! \downarrow,\uparrow \! \downarrow}$.
The functions
$\langle\langle X_p^{0, \alpha \! \uparrow\!}|
\!X_{p'}^{\alpha \! \uparrow,0}\rangle\rangle$ 
and
$\langle\langle X_p^{0, \alpha \! \uparrow\!}|\!Y^+_{p'}\rangle\rangle$ 
satisfy the equations:
\begin{eqnarray} \label{sys1}
&&(E+\tilde{\mu})\langle\langle X_p^{0, \alpha \! \uparrow\!}|\!X_{p'}^{\alpha \! \uparrow,0}\rangle\rangle=
{\langle X_p^{0}+X_{p'}^{\alpha \! \uparrow}\rangle \over 2\pi}\delta_{ij}+
\langle\langle [X_p^{0, \alpha \! \uparrow\!};H_{1}+H_{2}]|\!X_{p'}^{\alpha \! \uparrow,0}\rangle\rangle,
\nonumber\\
&&(E+\tilde{\mu}-U)\langle\langle Y_{p}|\!X_{p'}^{\alpha \! \uparrow, 0}\rangle\rangle=
\langle\langle [Y_{p};H_{1}+H_{2}]|\!X_{p'}^{\alpha \! \uparrow, 0}\rangle\rangle.
\end{eqnarray}
For calculation of these functions we use the generalized mean-field 
approximation~\cite{2_19}. Let us take the commutators in (\ref{sys1})
in the form:
\begin{eqnarray} \label{GHFA1}
\left[X_p^{0,\alpha\uparrow},H_{1}\right]&=&{\sum
\limits_j}'\epsilon (pj) X_j^{0,\alpha\uparrow},\ \
\left[X_p^{0,\alpha\uparrow},H_{2} \right]={\sum
\limits_j}'\tilde{\epsilon}(pj) Y_j,
\nonumber\\
\left[Y_p,H_{1}\right]&=&{\sum\limits_j}'
\zeta (pj) Y_j,\ \
\left[Y_p,H_{2} \right]={\sum
\limits_j}'\tilde{\zeta}(pj) X_j^{0,\alpha\uparrow},
\end{eqnarray}
where
$\epsilon(pj),\tilde{\epsilon}(pj),\zeta(pj),\tilde{\zeta}(pj)$ 
are the non-operator expressions.

After transition to ${\bf k}$-representation the system of 
equations (\ref{sys1}) taking into account (\ref{GHFA1}) has the 
solutions:
\begin{eqnarray} \label{1}
&&\langle\langle X_p^{0, \alpha \! \uparrow\!}|\!X_{p'}^{\alpha \! \uparrow,0}\rangle\rangle_{\bf{k}}=
{\langle X_p^{0}
+X_{p}^{\alpha \! \uparrow}\rangle \over 2\pi}
{E+\tilde{\mu}-U-\zeta({\bf k})
\over {(E-E_{1}({\bf k}))(E-E_{2}({\bf k}))}},
\nonumber\\
&&\langle\langle Y_{p}|\!X_{p'}^{\alpha \! \uparrow, 0}\rangle\rangle_{\bf{k}}=
{\langle X_p^{0}
+X_{p}^{\alpha \! \uparrow}\rangle \over 2\pi}
{\tilde{\zeta}({\bf k})
\over {(E-E_{1}({\bf k}))(E-E_{2}({\bf k}))}}.
\end{eqnarray}
Here $E_{1,2}({\bf k})$ is the quasiparticle energy spectrum:
\begin{eqnarray} \label{spectr1}
E_{1,2}({\bf k})=-\tilde{\mu}+{U \over 2}+{\epsilon({\bf k})+\zeta({\bf k}) \over 2}
\mp{1\over 2} \left\{ [U-\epsilon({\bf k})+\zeta({\bf k})]^2
+4\tilde{\epsilon}({\bf k})\tilde{\zeta}({\bf k}) \right\}^{1/2}.
\end{eqnarray}

 The same expression for the quasiparticle energy spectrum we obtain
from the functions
$\langle\langle X_p^{0, \alpha \! \uparrow\!}|\!Y^{+}_{p'}\rangle\rangle_{\bf{k}}$ 
and $\langle\langle Y_p|\!Y^{+}_{p'}\rangle\rangle_{\bf{k}}$.
The values of 
$\epsilon(pj),\tilde{\epsilon}(pj),\zeta(pj),\tilde{\zeta}(pj)$ 
we find by anticommutation of (\ref{GHFA1}) with the operators
$X_{p'}^{\alpha \! \uparrow, 0\!}$ and $Y^{+}_{p'}$:
\begin{eqnarray}
(X_p^{0}+X_{p}^{\alpha\uparrow})\epsilon(pp')&=&
\left\{X_{p'}^{\alpha\uparrow, 0\!};
\left[X_p^{0,\alpha\uparrow},H_{1}\right]\right\},
\nonumber\\
(X_p^{\alpha \! \downarrow}
+X_{p}^{\beta \! \uparrow}
+X_{p}^{\beta \! \downarrow}
+X_{p}^{\uparrow \! \uparrow}
+X_{p}^{\uparrow \! \downarrow}
+X_{p}^{\alpha2})
\tilde{\epsilon}(pp')&=&\left\{Y^{+}_{p'};
\left[X_p^{0,\alpha\uparrow},H_{2}\right]\right\},
\\
(X_p^{\alpha \! \downarrow}
+X_{p}^{\beta \! \uparrow}
+X_{p}^{\beta \! \downarrow}
+X_{p}^{\uparrow \! \uparrow}
+X_{p}^{\uparrow \! \downarrow}
+X_{p}^{\alpha2})
\zeta(pp')&=&\left\{Y^{+}_{p'};
\left[Y_p,H_{1}\right]\right\},
\nonumber\\
(X_p^{0}+X_{p}^{\alpha\uparrow})
\tilde{\zeta}(pp')&=&\left\{X_{p'}^{\alpha\uparrow, 0\!};
\left[Y_p,H_{2}\right]\right\}.
\nonumber
\end {eqnarray}
By use of the mean-field approximation 
analogously to the above, in the case of $t'_{k}=t''_{k}$ we obtain
\begin{eqnarray}
&&\epsilon({\bf k})=t_{{\bf k}} \left((c+b)+{3b^2 \over c+b} \right)-
3\tilde{t}_{{\bf k}}{6d^2 \over c+b},
\nonumber\\
&&\tilde{\epsilon}({\bf k})={{t}_{\bf k}+\tilde{t}_{\bf k} \over 2}
\left((c+b)-{b^2 \over d+b}+{6d^2 \over d+b}\right),
\\
&&\zeta({\bf k})=-t_{\bf k}{6d^2 \over d+b}+
\tilde{t}_{{\bf k}}\left(3(d+b)+{b^2 \over d+b}+{8d^2 \over 3(d+b)} \right),
\nonumber\\
&&\tilde{\zeta}({\bf k})={{t}_{\bf k}+\tilde{t}_{\bf k} \over 2}
\left((d+b)-{b^2 \over c+b}+{6d^2 \over c+b}\right),
\nonumber
\end{eqnarray}
here $c,\ b,\ d$ are the concentrations of the holes and sites occupied by 
one, two electrons, respectively, connected by the relations: 
\begin{eqnarray}
c=6d,\ \
b={1 \over 4}-3d;
\end{eqnarray}
and
\begin{eqnarray}
&&\tilde{t}_{\bf k}=t_{\bf k}+2t'_{\bf k}.
\end{eqnarray}
In the point of transition, when the concentrations of the holes and doublons
are equal to zero, the energies of the electrons within the subbands are
\begin{eqnarray}
&&E_{1}({\bf k})=-\tilde{\mu}+t_{\bf k},
\nonumber\\
&&E_{2}({\bf k})=-\tilde{\mu}+U+\tilde{t}_{\bf k}.
\end{eqnarray}
The energy gap in this case is
\begin{eqnarray} \label{gap1}
\Delta E=U-w-\tilde{w}=0,
\end{eqnarray} 
where $w=z|t_{ij}|$, $\tilde{w}=z|\tilde{t}_{ij}|$.
From the equation~(\ref{gap1}) we obtain the criterion of MIT:
\begin{eqnarray} \label{crt1}
U_á=w+\tilde{w}.
\end{eqnarray} 
In the partial case $t'_{\bf k}=t''_{\bf k}=0$ (in this case
$t_{\bf k}=\tilde{t}_{\bf k}$) 
we have
\begin{eqnarray} \label{crit1}
{U_á \over 2w}=1.
\end{eqnarray} 
In the Fig.~\ref{deg1} the energy gap for different values of the correlated hopping 
at $T=0$ is plotted. With the increase of the correlated hopping at the fixed
value of parameter $U/2w$ the energy gap width increases and the region
of values of $U/2w$ at which the system is in a metallic state, decreases. 

\subsection{The limit of the weak Hund's coupling at electron concentration
$n=2$}

Let us consider the MIT at electron concentration $n=2$.
In the vicinity of the transition point in the case of two electrons per 
atom the concentrations of holes and sites occupied by four electrons
are small. Neglecting the small amounts of these sites we can write 
the electron operators in the form:
\begin{eqnarray}
&&a_{i\alpha\uparrow}^{+}=X_i^{\uparrow\uparrow,\beta\uparrow}
			+X_i^{\uparrow\downarrow,\beta\downarrow}
			+X_i^{\alpha2,\alpha\downarrow}
			+X_i^{\alpha2\downarrow,\downarrow\downarrow}
			+X_i^{\alpha2\uparrow,\downarrow\uparrow}
			+X_i^{\beta2\uparrow,\beta2}, \ \
\nonumber\\ 
&&a_{i\alpha\downarrow}^{+}=X_i^{\downarrow\downarrow,\beta\downarrow}
			+X_i^{\downarrow\uparrow,\beta\uparrow}
			-X_i^{\alpha2,\alpha\uparrow}
			-X_i^{\alpha2\downarrow,\uparrow\downarrow}
			-X_i^{\alpha2\uparrow,\uparrow\uparrow}
			+X_i^{\beta2\downarrow,\beta2}, \ \
\\ 
&&a_{i\beta\uparrow}^{+}=-X_i^{\uparrow\uparrow,\alpha\uparrow}
			-X_i^{\downarrow\uparrow,\alpha\downarrow}
			+X_i^{\beta2,\beta\downarrow}
			-X_i^{\beta2\downarrow,\downarrow\downarrow}
			-X_i^{\beta2\uparrow,\uparrow\downarrow}
			+X_i^{\alpha2\uparrow,\alpha2}, \ \
\nonumber \\
&&a_{i\beta\downarrow}^{+}=-X_i^{\downarrow\downarrow,\alpha\downarrow}
			-X_i^{\uparrow\downarrow,\alpha\uparrow}
			-X_i^{\beta2,\beta\uparrow}
			+X_i^{\alpha2\downarrow,\alpha2}
			+X_i^{\beta2\uparrow,\uparrow\uparrow}
			+X_i^{\beta2\downarrow,\downarrow\uparrow}. \ \
\nonumber
\end{eqnarray}
Let us rewrite the Hamiltonian (\ref{H3}) in the configurational 
representation at electron concentration $n=2$. 
For the small values of the intra-atomic exchange interaction ($J \ll U$)
we take $J$ into account in the mean-field approximation (\ref{HF}). 
The Hamiltonian takes the form:
\begin{eqnarray}
H&=&-\tilde{\mu} \left(\sum_{i\gamma\sigma}X_i^{\gamma\sigma}
+2 \sum_{i\sigma}(X_i^{\sigma\sigma}+X_i^{\sigma\bar{\sigma}})
+2 \sum_{i\gamma}X_i^{\gamma2}
+3 \sum_{i\gamma\sigma} X_i^{\gamma2 \sigma} \right)
\nonumber \\
&+&U\left(\sum_{i\sigma}X_i^{\sigma\sigma}
+\sum_{i\sigma}X_i^{\sigma\bar{\sigma}}
+\sum_{i\gamma}X_i^{\gamma2}
+3 \sum_{i\gamma\sigma} X_i^{\gamma2 \sigma}\right)
+H_1+H_2,
\end{eqnarray} 
where the kinetic part of the Hamiltonian is
\begin{eqnarray}
&&H_1=\sum_{i,j}\Biggl(
(t_{ij}+2t''_{ij})\sum_{\gamma \sigma}X_i^{\gamma2,\gamma\sigma}
X_j^{\gamma\sigma,\gamma2}
\nonumber\\
&&+(t_{ij}+2t'_{ij})\biggl(\sum_{\gamma \sigma}
X_i^{\sigma \sigma,\gamma \sigma}
X_j^{\gamma \sigma,\sigma \sigma}
+\sum_{\sigma}\Bigl(X_i^{\sigma {\bar\sigma},\alpha \sigma}
X_j^{\alpha \sigma,\sigma {\bar\sigma}}
\nonumber\\
&&+X_i^{\sigma {\bar\sigma},\beta {\bar \sigma}}
X_j^{\beta {\bar \sigma},\sigma {\bar\sigma}}
+(X_i^{\sigma\sigma,\alpha\sigma}
X_j^{\alpha{\bar\sigma},{\bar\sigma} \sigma}
+X_i^{\sigma\sigma,\beta\sigma}
X_j^{\beta{\bar\sigma},\sigma{\bar\sigma}}+h.c.)\Bigr)\biggr)
\\
&&+(t_{ij}+t'_{ij}+t''_{ij})
\Bigl( \sum_{\gamma \sigma}\eta_{\sigma}\eta_{{\bar\gamma}}
X_i^{\sigma\sigma,\gamma\sigma}
X_j^{{\bar\gamma}{\bar\sigma},{\bar\gamma}2}
\nonumber\\
&&+\sum_{\sigma}\eta_{\sigma}
(X_i^{\sigma{\bar\sigma},\beta{\bar\sigma}}
X_j^{\alpha{\bar\sigma},\alpha2}+
X_i^{\sigma{\bar\sigma},\alpha\sigma}
X_j^{\beta\sigma,\beta2}) +h.c.\Bigr) 
\nonumber\\
&&+(t_{ij}+2t'_{ij}+2t''_{ij})\Bigl(
\sum_{\gamma\sigma}
X_i^{\gamma2\sigma,\sigma\sigma}
X_j^{\sigma\sigma,\gamma2\sigma}
+\sum_{\sigma}(X_i^{\alpha2\sigma,{\bar\sigma}\sigma}
X_j^{{\bar\sigma}\sigma,\alpha2\sigma}
\nonumber\\
&&+X_i^{\beta2\sigma,\sigma{\bar\sigma}}
X_j^{\sigma{\bar\sigma},\beta2\sigma})\Bigr)
+(t_{ij}+4t'_{ij})\sum_{\gamma\sigma}
X_i^{\gamma2 \sigma,\gamma2}
X_j^{\gamma2,\gamma2 \sigma}
\nonumber\\
&&+(t_{ij}+2t'_{ij}+2t''_{ij})
\sum_{\sigma}
(X_i^{\alpha2 \sigma,\sigma\sigma}
X_j^{\sigma{\bar\sigma},\alpha2 {\bar\sigma}}
+X_i^{\beta2 \sigma,\sigma\sigma}
X_j^{{\bar\sigma}\sigma,\beta2{\bar\sigma}}+h.c.)
\nonumber\\ 
&&+(t_{ij}+3t'_{ij}+t''_{ij})
\sum_{\gamma\sigma}(\eta_{\bar\sigma}\eta_{\gamma}
X_i^{\gamma2 \sigma,\sigma\sigma}
X_j^{{\bar\gamma2},{\bar\gamma2}{\bar\sigma}}+h.c.)
\nonumber\\ 
&&+(t_{ij}+3t'_{ij}+t''_{ij})
\sum_{\sigma}\eta_{\sigma}
(X_i^{\alpha2 \sigma,{\bar\sigma}\sigma}
X_j^{\beta2,\beta2 \sigma}
-X_i^{\beta2 \sigma,\sigma{\bar\sigma}}
X_j^{\alpha2,\alpha2\sigma}+h.c.)\Biggr), 
\nonumber
\end{eqnarray}
\begin{eqnarray} 
H_2=&&\sum_{i,j} \Biggl(
(t_{ij}+2t'_{ij}+t''_{ij})
\biggl(\sum_{\gamma\sigma}(\eta_{\sigma}
X_i^{\sigma\sigma,\gamma\sigma}
X_j^{{\bar\sigma}{\bar\sigma},{\bar\gamma}2{\bar\sigma}}
+\eta_{\bar\sigma}
X_i^{\gamma2,\gamma\sigma}X_j^{{\bar\gamma}2,{\bar\gamma}2{\bar\sigma}})
\nonumber\\
&&+\sum_{\sigma}\Bigl(\eta_{\sigma}(
X_i^{\sigma\sigma,\beta\sigma}
X_j^{{\bar\sigma}\sigma,\alpha2 \sigma}
+X_i^{\sigma\sigma,\alpha\sigma}
X_j^{\sigma{\bar\sigma},\beta2 \sigma}
+X_i^{\sigma{\bar\sigma},\beta {\bar\sigma}}
X_j^{{\bar\sigma}{\bar\sigma},\alpha2{\bar\sigma}}
\nonumber\\
&&+X_i^{\sigma{\bar\sigma},\beta {\bar\sigma}}
X_j^{{\bar\sigma}\sigma,\alpha2\sigma})
+\eta_{\bar\sigma}(
X_i^{\sigma{\bar\sigma},\alpha\sigma}
X_j^{{\bar\sigma}\sigma,\beta2{\bar\sigma}}-
X_i^{\sigma{\bar\sigma},\alpha{\bar\sigma}}
X_j^{{\bar\sigma}{\bar\sigma},\beta2{\bar\sigma}})\Bigr)\biggr)
\nonumber\\
&&+(t_{ij}+3t'_{ij})\Bigl(\sum_{\gamma\sigma}\eta_{\bar\sigma}
X_i^{\sigma\sigma,\gamma\sigma}
X_j^{\gamma2,\gamma2\sigma}
+\sum_{\sigma}
(X_i^{\sigma{\bar\sigma},\beta{\bar\sigma}}
X_j^{\beta2,\beta2\sigma}
\nonumber\\
&&-X_i^{\sigma{\bar\sigma},\alpha\sigma}
X_j^{\alpha2,\alpha2{\bar\sigma}})\Bigr)
+(t_{ij}+t'_{ij}+2t''_{ij})
\Bigl(\sum_{\gamma\sigma}\eta_{\gamma}
X_i^{\gamma2,\gamma\sigma}X_j^{\sigma\sigma,\gamma2\sigma}+
\nonumber\\
&&\sum_{\sigma}
(X_i^{\alpha2,\alpha \sigma}
X_j^{\sigma{\bar\sigma},\alpha2 {\bar\sigma}}
-X_i^{\beta2,\beta \sigma}
X_j^{{\bar\sigma}\sigma,\beta2{\bar\sigma}})\Bigr)
+h.c. \Biggr).
\nonumber     
\end{eqnarray}
The processes that form energy subbands are included in 
the Hamiltonian $H_1$, the processes  of 
hybridization of these subbands are included in the Hamiltonian $H_2$.

 Let us write the single-particle Green function as
\begin{eqnarray}
\langle\langle a_{p\alpha\uparrow}|a^{+}_{p'\alpha\uparrow}\rangle\rangle=
\langle\langle Y_p|\!Y^{+}_{p'}\rangle\rangle
+\langle\langle Z_p|\!Y^+_{p'}\rangle\rangle +\langle\langle Y_p|\!Z^{+}_{p'}\rangle\rangle
+\langle\langle Z_p|\!Z^+_{p'}\rangle\rangle.
\end{eqnarray}

Here the following notations have been introduced:  
$Y_p=X_{p}^{\alpha \! \downarrow,\alpha2}
+X_{p}^{\beta \! \uparrow,\uparrow \! \uparrow}
+X_{p}^{\beta \! \downarrow,\uparrow \! \downarrow}$,
$Z_p=X_{p}^{\downarrow \! \downarrow,\alpha2 \downarrow}
+X_{p}^{\downarrow \! \uparrow,\alpha2 \uparrow}
+X_{p}^{\beta2,\beta2 \uparrow}$.
The functions
$\langle\langle Y_p|\!Y^+_{p'}\rangle\rangle$ 
and $\langle\langle Z_p|\!Y^+_{p'}\rangle\rangle$ 
satisfy the equations:
\begin{eqnarray} \label{sys2}
&&(E+\tilde{\mu}-U)\langle\langle Y_p|\!Y^{+}_{p'}\rangle\rangle=
{A \over 2\pi}\delta_{ij}+
\langle\langle [Y_p;H_{1}+H_{2}]|\!Y^{+}_{p'}\rangle\rangle,
\nonumber\\
&&(E+\tilde{\mu}-2U)\langle\langle Z_{p}|\!Y^{+}_{p'}\rangle\rangle=
\langle\langle [Z_{p};H_{1}+H_{2}]|\!Y^{+}_{p'}\rangle\rangle,
\end{eqnarray}
where $A=\langle X_{p}^{\alpha \! \downarrow}
+X_{p}^{\beta \! \uparrow}
+X_{p}^{\beta \! \downarrow}
+X_{p}^{\uparrow \! \uparrow}
+X_{p}^{\uparrow \! \downarrow}
+X_{p}^{\alpha2} \rangle$.
On the analogy of the previous section we use the generalized mean-field
approximation to calculate these functions. Let us take the commutators 
in (\ref{sys2}) in the form:
\begin{eqnarray} \label{GHFA2}
\left[Y_p,H_{1}\right]&=&{\sum \limits_j}'\epsilon (pj) Y_j,\ \
\left[Y_p,H_{2} \right]={\sum \limits_j}'\tilde{\epsilon}(pj) Z_j,
\nonumber\\
\left[Z_p,H_{1}\right]&=&{\sum\limits_j}'\zeta (pj) Z_j,\ \
\left[Z_p,H_{2} \right]={\sum \limits_j}'\tilde{\zeta}(pj) Y_j,
\end{eqnarray}
where $\epsilon(pj),\tilde{\epsilon}(pj),\zeta(pj),\tilde{\zeta}(pj)$ 
are the non-operator expressions, which we calculate using 
method of paper~\cite{2_19}.  

After transition to ${\bf k}$-representation the system of 
equations (\ref{sys2}) taking into account (\ref{GHFA2}) has the 
solutions:
\begin{eqnarray} \label{solv2}
&&\langle\langle Y_p|\!Y^{+}_{p'}\rangle\rangle_{\bf{k}}=
{A \over 2\pi}{E+\tilde{\mu}-2U-\zeta({\bf k})
\over {(E-E_{1}({\bf k}))(E-E_{2}({\bf k}))}},
\nonumber\\
&&\langle\langle Z_{p}|\!Y^{+}_{p'}\rangle\rangle_{\bf{k}}=
{A \over 2\pi}{\tilde{\zeta}({\bf k})
\over {(E-E_{1}({\bf k}))(E-E_{2}({\bf k}))}}.
\end{eqnarray}

Here $E_{1,2}({\bf k})$ is the quasiparticle energy spectrum:
\begin{eqnarray} \label{spectr2}
E_{1,2}({\bf k})=-\tilde{\mu}+{3U \over 2}+{\epsilon({\bf k})+\zeta({\bf k}) \over 2}
\mp{1\over 2} \left\{ [U-\epsilon({\bf k})+\zeta({\bf k})]^2
+4\tilde{\epsilon}({\bf k})\tilde{\zeta}({\bf k}) \right\}^{1/2}.
\end{eqnarray}

 The same expression for the quasiparticle energy spectrum we obtain
from the functions
$\langle\langle Y_p|\!Z^{+}_{p'}\rangle\rangle_{\bf{k}}$ and
$\langle\langle Z_p|\!Z^{+}_{p'}\rangle\rangle_{\bf{k}}$.
The values of 
$\epsilon(pj),\tilde{\epsilon}(pj),\zeta(pj),\tilde{\zeta}(pj)$ 
we find by anticommutation of (\ref{GHFA2}) with the operators
$Y^{+}_{p'}$ and $Z^{+}_{p'}$:
\begin{eqnarray}
(X_p^{\alpha \! \downarrow}
+X_{p}^{\beta \! \uparrow}
+X_{p}^{\beta \! \downarrow}
+X_{p}^{\uparrow \! \uparrow}
+X_{p}^{\uparrow \! \downarrow}
+X_{p}^{\alpha2})\epsilon(pp')&=&
\left\{Y^{+}_{p'};\left[Y_p,H_{1}\right]\right\},
\nonumber\\
(X_p^{\downarrow \! \downarrow}
+X_{p}^{\downarrow \! \uparrow}
+X_{p}^{\beta2}
+X_{p}^{\alpha2 \uparrow}
+X_{p}^{\alpha2 \downarrow}
+X_{p}^{\beta2 \uparrow})
\tilde{\epsilon}(pp')&=&\left\{Z^{+}_{p'};
\left[Y_p,H_{2}\right]\right\},
\nonumber\\
(X_p^{\downarrow \! \downarrow}
+X_{p}^{\downarrow \! \uparrow}
+X_{p}^{\beta2}
+X_{p}^{\alpha2 \uparrow}
+X_{p}^{\alpha2 \downarrow}
+X_{p}^{\beta2 \uparrow})
\zeta(pp')&=&\left\{Z^{+}_{p'};
\left[Z_p,H_{1}\right]\right\},
\\
((X_p^{\alpha \! \downarrow}
+X_{p}^{\beta \! \uparrow}
+X_{p}^{\beta \! \downarrow}
+X_{p}^{\uparrow \! \uparrow}
+X_{p}^{\uparrow \! \downarrow}
+X_{p}^{\alpha2})
\tilde{\zeta}(pp')&=&\left\{Y^{+}_{p'};
\left[Z_p,H_{2}\right]\right\}.
\nonumber
\end {eqnarray}
By use of the mean-field approximation 
analogously to the above, in the case of $t'_{k}=t''_{k}$ we obtain
\begin{eqnarray}
&&\epsilon({\bf k})=\tilde{t}_{{\bf k}} \left(3(d+b)+{b^2 \over d+b} 
+{8d^2 \over 3(d+b)} \right)-t^{*}_{{\bf k}}{8b^2 \over 3(d+b)},
\nonumber\\
&&\tilde{\epsilon}({\bf k})={\tilde{t}_{\bf k}+t^{*}_{\bf k} \over 2}
\left(3(d+b)-{8d^2 \over 3(d+b)}+{7b^2 \over 3(d+b)}\right),
\\
&&\zeta({\bf k})=-\tilde{t}_{\bf k}{8b^2 \over 3(d+b)}+
t^{*}_{{\bf k}}\left(3(d+b)+{b^2 \over d+b}+{8d^2 \over 3(d+b)} \right),
\nonumber\\
&&\tilde{\zeta}({\bf k})={\tilde{t}_{\bf k}+t^{*}_{\bf k} \over 2}
\left(3(d+b)-{8d^2 \over 3(d+b)}+{7b^2 \over 3(d+b)}\right),
\nonumber
\end{eqnarray}
with
\begin{eqnarray}
&&\tilde{t}_{\bf k}=t_{\bf k}+2t'_{\bf k}.
\nonumber\\
&&t^{*}_{\bf k}=t_{\bf k}+4t'_{\bf k};
\end{eqnarray}
here $b$ is the concentration of the sites occupied by 
one (or three) electrons, $d$ is the concentration of the doubly occupied 
sites, connected by the relation: 
\begin{eqnarray}
b={1-8d \over 6}.
\end{eqnarray}

In the point of transition, when the concentrations of the singly and
triply occupied sites are equal to zero, the quasiparticle energy spectrum is
\begin{eqnarray} \label{spctr2}
E_{1,2}({\bf k})=-\tilde{\mu}+{3U \over 2}+{17 \over 18}
{t^{*}_{\bf k}+\tilde{t}_{\bf k} \over 2}
\mp{1\over 2} \left\{ \Bigl[U+{17 \over 18}
{t^{*}_{\bf k}-\tilde{t}_{\bf k} \over 2} \Bigr]^2
+\Bigl[{t^{*}_{\bf k}+\tilde{t}_{\bf k} \over 18} \Bigr]^2 \right\}^{1/2}.
\end{eqnarray}

Using the  quasiparticle energy spectrum~(\ref{spctr2})  we find the 
energy gap width. In the point of MIT the energy gap is equal to zero.
From this  condition we find the criterion of MIT. In the partial case of 
$t'_{\bf k}=t''_{\bf k}=0$ (in this case $t^{*}_{\bf k}=\tilde{t}_{\bf k}$)
we find
\begin{eqnarray} \label{crit2}
{U_á \over 2w}={2\sqrt{2} \over 3}.
\end{eqnarray} 
In the Fig.~\ref{deg2} the energy gap for different values of the correlated hopping 
at $T=0$ is plotted. With the increase of the correlated hopping at the fixed
value of parameter $U/2w$ the energy gap width increases faster than at
$n=1$ and the region of values of $U/2w$ at which the system 
is in the metallic state, decreases, analogously to the case $n=1$. 
\subsection{The limit of the weak Hund's coupling at electron concentration
$n=3$}

Let us consider the MIT at electron concentration $n=3$.
In the vicinity of the transition point in the case of three electrons per 
atom the concentrations of holes and sites occupied by one electron
are small. Neglecting the small amounts of these sites we can write 
the electron operators in the form:
\begin{eqnarray}
a_{i\alpha\uparrow}^{+}=+X_i^{\alpha2\downarrow,\downarrow\downarrow}
			+X_i^{\alpha2\uparrow,\downarrow\uparrow}
			+X_i^{\beta2\uparrow,\beta2}
			+X_i^{4,\beta2\downarrow}, \ \
\nonumber\\ 
a_{i\alpha\downarrow}^{+}=-X_i^{\alpha2\downarrow,\uparrow\downarrow}
			-X_i^{\alpha2\uparrow,\uparrow\uparrow}
			+X_i^{\beta2\downarrow,\beta2}
			-X_i^{4,\beta2\uparrow}, \ \
\\ 
a_{i\beta\uparrow}^{+}=-X_i^{\beta2\downarrow,\downarrow\downarrow}
			-X_i^{\beta2\uparrow,\uparrow\downarrow}
			+X_i^{\alpha2\uparrow,\alpha2}
			+X_i^{4,\alpha2\downarrow}, \ \
\nonumber \\
a_{i\beta\downarrow}^{+}=+X_i^{\alpha2\downarrow,\alpha2}
			+X_i^{\beta2\uparrow,\uparrow\uparrow}
			+X_i^{\beta2\downarrow,\downarrow\uparrow}
			-X_i^{4,\alpha2\uparrow}. \ \
\nonumber
\end{eqnarray}

Let us rewrite the Hamiltonian (\ref{H3}) in the configurational representation 
at electron concentration $n=3$. 
For the small values of the intra-atomic exchange interaction ($J \ll U$)
we take $J$ into account in the mean field approximation (\ref{HF}). 
The Hamiltonian takes the form:
\begin{eqnarray}
H&=&-\tilde{\mu} \left(
2 \sum_{i\sigma}(X_i^{\sigma\sigma}+X_i^{\sigma\bar{\sigma}})
+2 \sum_{i\gamma}X_i^{\gamma2}
+3 \sum_{i\gamma\sigma} X_i^{\gamma2 \sigma} 
+4 \sum_{i}X_i^{4}\right)
\nonumber \\
&+&U\left(\sum_{i\sigma}X_i^{\sigma\sigma}
+\sum_{i\sigma}X_i^{\sigma\bar{\sigma}}
+\sum_{i\gamma}X_i^{\gamma2}
+3 \sum_{i\gamma\sigma} X_i^{\gamma2 \sigma}
+6 X_i^{4}\right)
+H_1+H_2,
\end{eqnarray} 
where the kinetic part of the Hamiltonian 
\begin{eqnarray}
&&H_1=\sum_{i,j}\Biggl(
(t_{ij}+2t'_{ij}+2t''_{ij})\Bigl(
\sum_{\gamma\sigma}
X_i^{\gamma2\sigma,\sigma\sigma}
X_j^{\sigma\sigma,\gamma2\sigma}
+\sum_{\sigma}(X_i^{\alpha2\sigma,{\bar\sigma}\sigma}
X_j^{{\bar\sigma}\sigma,\alpha2\sigma}
\nonumber\\
&&+X_i^{\beta2\sigma,\sigma{\bar\sigma}}
X_j^{\sigma{\bar\sigma},\beta2\sigma}+h.c.)\Bigr)
+(t_{ij}+4t'_{ij})\sum_{\gamma\sigma}
X_i^{\gamma2 \sigma,\gamma2}
X_j^{\gamma2,\gamma2 \sigma}
\nonumber\\
&&+(t_{ij}+2t'_{ij}+2t''_{ij})
\sum_{\sigma}
(X_i^{\alpha2 \sigma,\sigma\sigma}
X_j^{\sigma{\bar\sigma},\alpha2 {\bar\sigma}}
+X_i^{\beta2 \sigma,\sigma\sigma}
X_j^{{\bar\sigma}\sigma,\beta2{\bar\sigma}}+h.c.)
\nonumber\\ 
&&+(t_{ij}+3t'_{ij}+t''_{ij})
\sum_{\gamma\sigma}(\eta_{\bar\sigma}\eta_{\gamma}
X_i^{\gamma2 \sigma,\sigma\sigma}
X_j^{{\bar\gamma2},{\bar\gamma2}{\bar\sigma}}+h.c.)
\nonumber\\ 
&&+(t_{ij}+3t'_{ij}+t''_{ij})
\sum_{\sigma}\eta_{\sigma}
(X_i^{\alpha2 \sigma,{\bar\sigma}\sigma}
X_j^{\beta2,\beta2 \sigma}
-X_i^{\beta2 \sigma,\sigma{\bar\sigma}}
X_j^{\alpha2,\alpha2\sigma}+h.c.)
\nonumber\\
&&+(t_{ij}+4t'_{ij}+2t''_{ij})
X_i^{4,\gamma2 \sigma}
X_j^{\gamma2 \sigma,4} \Biggr),
\nonumber
\end{eqnarray}
\begin{eqnarray} 
H_2=&&\sum_{i,j} \Biggl(
(t_{ij}+3t'_{ij}+2t''_{ij})
\sum_{\gamma\sigma} \biggl( \eta_{\gamma}
X_i^{\gamma2\sigma,\sigma\sigma}
X_j^{{\bar\gamma}2\sigma,4}
\nonumber\\
&&+\sum_{\sigma}\Bigl(
X_i^{\alpha 2 \sigma,{\bar\sigma}\sigma}
X_j^{\beta 2{\bar\sigma},4}
-X_i^{\beta 2 \sigma,\sigma{\bar\sigma}}
X_j^{\alpha 2{\bar\sigma},4} \Bigr) 
\biggr)
\nonumber\\
&&+(t_{ij}+3t'_{ij})\sum_{\gamma\sigma}\eta_{\sigma}
X_i^{\gamma2\sigma,\gamma2}
X_j^{\gamma2{\bar\sigma},4}
+h.c. \Biggr).
\nonumber     
\end{eqnarray}
The processes that form energy subbands are included in the 
Hamiltonian $H_1$, the processes  of 
hybridization of these subbands are included in the Hamiltonian $H_2$.

 Let us write the single-particle Green function as:
\begin{eqnarray}
\langle\langle a_{p\alpha\uparrow}|a^{+}_{p'\alpha\uparrow}\rangle\rangle=
\langle\langle X_p^{\beta 2\! \downarrow\!,4}|\!
X_{p'}^{4, \beta 2\! \downarrow\!}\rangle\rangle
+\langle\langle X_p^{\beta 2\! \downarrow\!,4}|\!
Z^+_{p'}\rangle\rangle \\
+\langle\langle Z_p|\!X_{p'}^{4, \beta 2\! \downarrow\!}\rangle\rangle
+\langle\langle Z_p|\!Z^+_{p'}\rangle\rangle.
\nonumber
\end{eqnarray}

Here the following notations have been introduced:  
$Z_p=X_{p}^{\downarrow \! \downarrow,\alpha2 \downarrow}
+X_{p}^{\downarrow \! \uparrow,\alpha2 \uparrow}
+X_{p}^{\beta2,\beta2 \uparrow}$.
The functions $\langle\langle X_p^{\beta 2\! \downarrow\!,4}|\!
X_{p'}^{4, \beta 2\! \downarrow\!}\rangle\rangle$ and
$\langle\langle Z_p|\!X_{p'}^{4, \beta 2\! \downarrow\!}\rangle\rangle$
satisfy the equations:
\begin{eqnarray} \label{sys3}
&&\hspace{-10mm}(E+\tilde{\mu}-3U)\langle\langle X_p^{\beta 2\! \downarrow\!,4}|\!
X_{p'}^{4, \beta 2\! \downarrow\!}\rangle\rangle=
{\langle X_p^{\beta 2\! \downarrow\!}+X_{p}^{4}\rangle \over 2\pi}\delta_{ij}+
\langle\langle [X_p^{\beta 2\! \downarrow\!,4};H_{1}+H_{2}]|\!
X_{p'}^{4, \beta 2\! \downarrow\!}\rangle\rangle,
\nonumber\\
&&\hspace{-10mm}(E+\tilde{\mu}-2U)\langle\langle Z_{p}|\!
X_{p'}^{4, \beta 2\! \downarrow\!}\rangle\rangle=
\langle\langle [Z_{p};H_{1}+H_{2}]|\!X_p^{\alpha \! \uparrow, 0}\rangle\rangle.
\end{eqnarray}
On the analogy of the previous sections we use the generalized mean-field
approximation to calculate these functions. 
Let us take the commutators in (\ref{sys3}) in the form:
\begin{eqnarray} \label{GHFA3}
\left[X_p^{\beta 2\! \downarrow\!,4},H_{1}\right]&=&{\sum
\limits_j}'\epsilon (pj) X_j^{\beta 2\! \downarrow\!,4},\ \
\left[X_p^{\beta 2\! \downarrow\!,4},H_{2} \right]={\sum
\limits_j}'\tilde{\epsilon}(pj) Z_j,
\nonumber\\
\left[Z_p,H_{1}\right]&=&{\sum\limits_j}'
\zeta (pj) Z_j,\ \
\left[Z_p,H_{2} \right]={\sum
\limits_j}'\tilde{\zeta}(pj) X_j^{\beta 2\! \downarrow\!,4},
\end{eqnarray}
where $\epsilon(pj),\tilde{\epsilon}(pj),\zeta(pj),\tilde{\zeta}(pj)$ 
are the non-operator expressions.

After transition to ${\bf k}$-representation the system of 
equations (\ref{sys3}) taking into account (\ref{GHFA3}) has the 
solutions:
\begin{eqnarray} \label{3}
&&\langle\langle X_p^{\beta 2\! \downarrow\!,4}|\!
X_{p'}^{4, \beta 2\! \downarrow\!}\rangle\rangle_{\bf{k}}=
{\langle X_p^{\beta 2\! \downarrow\!}+X_{p}^{4}\rangle \over 2\pi}
{E+\tilde{\mu}-2U-\zeta({\bf k})
\over {(E-E_{1}({\bf k}))(E-E_{2}({\bf k}))}},
\nonumber\\
&&
\langle\langle Z_p|\!X_{p'}^{4, \beta 2\! \downarrow\!}\rangle\rangle_{\bf{k}}=
{\langle X_p^{\beta 2\! \downarrow\!}
+X_{p}^{4}\rangle \over 2\pi}{\tilde{\zeta}({\bf k})
\over {(E-E_{1}({\bf k}))(E-E_{2}({\bf k}))}}.
\end{eqnarray}

Here $E_{1,2}({\bf k})$ is the quasiparticle energy spectrum: 
\begin{eqnarray} \label{spectr3}
E_{1,2}({\bf k})=-\tilde{\mu}+{5U \over 2}+{\epsilon({\bf k})+\zeta({\bf k}) \over 2}
\mp{1\over 2} \left\{ [U-\epsilon({\bf k})+\zeta({\bf k})]^2
+4\tilde{\epsilon}({\bf k})\tilde{\zeta}({\bf k}) \right\}^{1/2}.
\end{eqnarray}

The same expression for the quasiparticle energy spectrum we obtain
from the functions
$\langle\langle X_p^{\beta 2\! \downarrow\!,4}|\!
Z^+_{p'}\rangle\rangle,\ 
\langle\langle Z_p|\!Z^+_{p'}\rangle\rangle$.

The values of 
$\epsilon(pj),\tilde{\epsilon}(pj),\zeta(pj),\tilde{\zeta}(pj)$ 
we find by anticommutation of (\ref{GHFA3}) with the operators
$X_{p'}^{4, \beta 2\! \downarrow\!}$ and $Z^{+}_{p'}$:
\begin{eqnarray}
(X_p^{\beta 2\! \downarrow\!}+X_{p}^{4})\epsilon(pp')&=&
\left\{X_{p'}^{4, \beta 2\! \downarrow\!};
\left[X_p^{\beta 2\! \downarrow\!,4},H_{1}\right]\right\},
\nonumber\\
(X_p^{\downarrow \! \downarrow}
+X_{p}^{\downarrow \! \uparrow}
+X_{p}^{\beta2}
+X_{p}^{\alpha2 \uparrow}
+X_{p}^{\alpha2 \downarrow}
+X_{p}^{\beta2 \uparrow})
\tilde{\epsilon}(pp')&=&\left\{Z^{+}_{p'};
\left[X_p^{0,\alpha\uparrow},H_{2}\right]\right\},
\\
(X_p^{\downarrow \! \downarrow}
+X_{p}^{\downarrow \! \uparrow}
+X_{p}^{\beta2}
+X_{p}^{\alpha2 \uparrow}
+X_{p}^{\alpha2 \downarrow}
+X_{p}^{\beta2 \uparrow})
\zeta(pp')&=&\left\{Z^{+}_{p'};
\left[Z_p,H_{1}\right]\right\},
\nonumber\\
(X_p^{\beta 2\! \downarrow\!}+X_{p}^{4})
\tilde{\zeta}(pp')&=&\left\{X_{p'}^{\alpha\uparrow, 0\!};
\left[Z_p,H_{2}\right]\right\}.
\nonumber
\end {eqnarray}
By use of the mean-field approximation 
analogously to the above, in the case $t'_{k}=t''_{k}$ we obtain
\begin{eqnarray}
&&\epsilon({\bf k})=t^{\bullet}_{{\bf k}} \left((t+f)+{3t^2 \over t+f} \right)-
3t^{*}_{{\bf k}}{3df \over t+f},
\nonumber\\
&&\tilde{\epsilon}({\bf k})={{t^{*}}_{\bf k}+{t^{\bullet}}_{\bf k} \over 2}
\left((t+f)-{t^2 \over d+t}+{df \over d+t}\right),
\\
&&\zeta({\bf k})=-t^{\bullet}_{\bf k}{df \over d+t}+
t^{*}_{{\bf k}}\left(3(d+t)+{t^2 \over d+t}+{4d^2 \over (d+t)} \right),
\nonumber\\
&&\tilde{\zeta}({\bf k})={{t^{*}}_{\bf k}+{t^{\bullet}}_{\bf k} \over 2}
\left((d+t)-{t^2 \over t+f}+{3df \over t+f}\right),
\nonumber
\end{eqnarray}
here 
\begin{eqnarray}
&&t^{*}_{\bf k}=t_{\bf k}+4t'_{\bf k},
\nonumber\\
&&t^{\bullet}_{\bf k}=t_{\bf k}+6t'_{\bf k};
\end{eqnarray}
$d,\ t,\ f$  are the concentrations of the sites occupied by 
two, three and four electrons, respectively, connected by the relations: 
\begin{eqnarray}
f=6d,\ \
t={1 \over 4}-3d.
\end{eqnarray}
  
In the point of transition, when the concentrations of the holes and single 
electrons are equal to zero, the energies of the electrons within 
the subbands are
\begin{eqnarray}
&&E_{1}({\bf k})=-\tilde{\mu}+2U+t^{*}_{\bf k},
\nonumber\\
&&E_{2}({\bf k})=-\tilde{\mu}+3U+t^{\bullet}_{\bf k}.
\end{eqnarray}
The energy gap in this case is
\begin{eqnarray} \label{gap3}
\Delta E=U-w^{*}-w^{\bullet}=0,
\end{eqnarray} 
where $w^{*}=z|t^{*}_{ij}|$, $w^{\bullet}=z|t^{\bullet}_{ij}|$.

From the equation~(\ref{gap3}) we obtain the criterion of the MIT
at the electron concentration $n=3$:
\begin{eqnarray} \label{crt3}
U_á=w^{*}+w^{\bullet}.
\end{eqnarray} 
In the partial case $t'_{\bf k}=t''_{\bf k}=0$ (in this case
$t_{\bf k}=\tilde{t}_{\bf k}$) 
we have
\begin{eqnarray} \label{crit3}
{U_á \over 2w}=1.
\end{eqnarray}
This result coincides with the corresponding critical value at the 
electron concentration $n=1$ in the consequence of the electron-hole
symmetry of the model without the correlated hopping.

In Fig.~\ref{de3} the energy gap for different values of the correlated hopping 
at $T=0$ is plotted. With the increase of the correlated hopping at the fixed
value of parameter $U/2w$ the energy gap width increases faster than at 
$n=1, n=2$ and the region of values of $U/2w$ at which the system is in 
a metallic state, decreases. 

\section{Discussions and conclusions}
\label{sec: concl}

In the present paper we have proposed a doubly orbitally degenerate 
narrow-band model with correlated hopping. The peculiarity of the
model is taking into account the matrix element of electron-electron
interaction which describes intersite hoppings of electrons. In particular,
this leads to the concentration dependence of the hopping 
integrals. Using the representation of Hamiltonian of a 
doubly orbitally degenerate model with correlated hopping in terms 
of the Hubbard operators the cases of the strong and weak 
Hund's coupling have been considered. By means 
of a generalized mean-field approximation we have calculated the 
single-particle Green function and quasiparticle energy spectrum. 
Metal-insulator transition has been studied in the model
at different integer values of the electron concentration. With the help
of the obtained energy spectrum we have found energy gap width and criteria 
of metal-insulator transition. 

The peculiarities of the expressions for quasiparticle energy spectrum
and energy gap are dependences on
the concentration of polar states (holes, doublons at $n=1$; single electron
and triple occupied sites at $n=2$; doublons and sites occupied by four
electrons at $n=3$), 
on the hopping integrals (thus on external pressure). At given values of 
$U$ and hopping integrals (constant external pressure) the concentration 
dependence of $\Delta E$ allows to study MIT under the action of external
influences. In particular,
$\Delta E(T)$-dependence can lead to the transition from a metallic state to
an insulating state with the increase of temperature (see Fig.~\ref{gap_t});
the described transition is observed, in particular,
in the (V$_{1-x}$Cr$_x$)$_2$O$_3$ compound~\cite{1_69,mcwhan}
and the NiS$_{2-x}$Se$_x$ system~\cite{wilson,honig}.
The similar dependence of energy gap width can be observed at change
of the polar states concentration under the action of photoeffect or 
magnetic field. The strong magnetic field can lead, for example, to the 
decrease of polar state concentration (see Ref.~\cite{1_25}) initiating 
the transition from
a paramagnetic insulator state to a paramagnetic metal state. 
Contrariwise, the increase of polar state concentration under the 
action of light stimulates the metal-insulator transition,
analogously to the influence of temperature change.
At the increase of bandwidth (for example, under the action of external 
pressure or composition changes) the insulator-to-metal 
transition can occur.

The results allow to study the influence of  the correlated hopping 
and orbital degeneracy on MIT. 
The dependences of energy gap width on the parameter $U/2w$ in absence
of the correlated hopping ($t'_{\bf k}=t''_{\bf k}=0$) at different 
electron concentrations are given in Fig.~\ref{deg1_2}. One can see that in the case
$n=2$ the MIT occurs at smaller value of $U/2w$ then at $n=1$.
This result is in qualitative accordance with the results of work~\cite{5},
in distinction from~\cite{7,9}.
Using the critical values of the papameter $U/(2w)$ at which MIT occurs for 
different integer electron concentrations (see Fig.~\ref{diag_int}) we can interpret 
the fact that in the series of 
disulphides MS$_2$ the CoS$_2$ (one electron within $e_g$ band corresponding
to $n=1$) and CuS$_2$ compounds (three electrons within $e_g$-band 
corresponding $n=3$) are metals, and the NiS$_2$ compound (two electrons
within $e_g$-band corresponding $n=2$) is an insulator.
Really, for $0.94\le U/2w\le 1$ at the electron concentration $n=2$
system described by the present model is an insulator, whereas for the same
values of the parameter $U/2w$ at the electron concentrations $n=1,\ 3$
system is a metal (according with the calculations of Ref.~\cite{bocq} the
ratios $U/2w$ in these compounds have close values). 

We have found that in the case of the strong Hund's coupling at $n=1$ 
metal-insulator transition occurs at smaller value of the parameter
$((U-J)/2w)_c=0.75$ than in the case of the weak Hund's coupling
$((U-J)/2w)_c=1$.

When the magnetically ordered states are taken into account the phase 
diagram of the considered model (Fig.~\ref{diag_int}) can be changed.
In particular, with increase of correlation strength the transition 
from paramagnetic to magnetically ordered state~\cite{3,4} (antiferromagnetic
insulator or ferromagnetic insulator) can occur, similarly 
to the magnetic transition found in the Ref.~\cite{6} by use of slave-boson 
method for the doubly degenerate Hubbard model. 

At nonzero values of correlated hopping the point of MIT moves towards the
values of parameter $U/2w$ (Figs.~\ref{deg1}-\ref{de3}) at which 
system is a metal in proportion to correlated hopping value 
(Fig.~\ref{crit123}). From Fig.~\ref{crit123} one can see that 
$(U/2w)_c$ decreases
with increasing correlated hopping, and what is more at $n=2$ 
with the increase of the correlated hopping parameter $\kappa$ the value 
$(U/2w)_c$ decreases faster than at the electron concentration $n=1$, and 
at $n=3$  the value $(U/2w)_c$ decreases faster than at $n=1,\ n=2$. The 
non-equivalence of the cases $n=1$ and $n=3$ is a manifestation of the
electron-hole asymmetry which is a characteristic of the models with 
correlated hopping. 

Thus both orbital degeneracy and correlated hopping are the factors
favoring the transition of system to an insulating state in the case of 
half-filling with the increase of intra-atomic Coulomb repulsion
in comparison with the single-band Hubbard model (in this connection
see Refs.~\cite{1_59,1_62}).

In the present paper considering MIT we have neglected
the correlated hopping $T_1$. Taking into account $T_1$ leads to the
concentration dependence of the hopping integrals and as a result to
decreasing $U_c$. This effect shows
itself the more the larger is the value of electron concentration $n$. 
A more detailed analysis of the correlated hopping $T_1$ influence will
be given in subsequent papers. 

\acknowledgments
The authors (Yu.S. and V.H.) are grateful to the Max Planck Institute for
the Physics of Complex Systems (Dresden, Germany) for the hospitality 
during the international workshop and seminar on ``Electronic and magnetic
properties of novel transition metal compounds: From cuprates to titanates''
(October 5-31, 1998) where the part of the results considered in the 
present paper were discussed.

\begin{figure}
\caption{The possible configurations of the lattice sites.}
\label{sites}
\end{figure}

\begin{figure}
\caption{Energy levels corresponding to the possible electron 
configurations of sites and the transitions between them.}
\label{levels}
\end{figure}

\begin{figure}
\caption{The dependence of energy gap width $\Delta E/(U'-J)$ 
on the parameter $(U'-J)/(2w)$: the upper curve --
$(kT)/(2w)=0.1$; the middle curve -- $(kT)/(2w)=0.05$; the lower curve --
$(kT)/(2w)=0$.}
\label{gap_w}
\end{figure}

\begin{figure}
\caption{The dependence of energy gap width $\Delta E/(U'-J)$ 
on the parameter $(kT)/(2w)$: the upper curve --
$(U'-J)/(2w)=0.74$;  the lower curve -- $(U'-J)/(2w)=0.72$.}
\label{gap_t}
\end{figure}

\begin{figure}
\caption{The dependence of energy gap width $\Delta E/U$ on the 
parameter $U/(2w)$ for $n=1$ at different values of the parameter
$\kappa=t'_{ij}/t_{ij}$: 
the lower curve --$\kappa=0$; the middle curve --$\kappa=0.2$;
the upper curve --$\kappa=0.6$.}
\label{deg1}
\end{figure}

\begin{figure}
\caption{The dependence of energy gap width $\Delta E/U$ on the 
parameter $U/(2w)$ for $n=2$ at different values of the parameter
$\kappa=t'_{ij}/t_{ij}$: 
the lower curve --$\kappa=0$; the middle curve --$\kappa=0.1$;
the upper curve --$\kappa=0.2$.}
\label{deg2}
\end{figure}

\begin{figure}
\caption{The dependence of energy gap width $\Delta E/U$ on the 
parameter $U/(2w)$ for $n=3$ at different values of the parameter
$\kappa=t'_{ij}/t_{ij}$: 
the lower curve --$\kappa=0$; the middle curve --$\kappa=0.1$;
the upper curve --$\kappa=0.15$.}
\label{de3}
\end{figure}

\begin{figure}
\caption{The dependence of energy gap width $\Delta E/U$ on the 
parameter $U/(2w)$ in the absence of correlated hopping
($t'_{\bf k}=t''_{\bf k}=0$):
the lower curve --$n=1$, $n=3$; the upper curve --$n=2$.}
\label{deg1_2}
\end{figure}

\begin{figure}
\caption{The electron vs. interaction phase diagram showing the 
paramagnetic metal (PM) and paramagnetic insulator (PI) in absence of
correlated hopping.}
\label{diag_int}
\end{figure}

\begin{figure}
\caption{The dependence of critical value $(U/2w)_c$ on the  
parameter of correlated hopping $\kappa=t'_{ij}/t_{ij}$: 
the curve 1 -  $n=1$; the curve 2 -  $n=2$; the curve 3 - $n=3$.}
\label{crit123}
\end{figure}

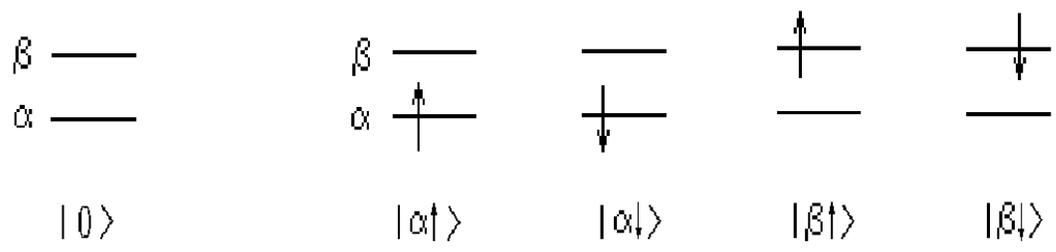
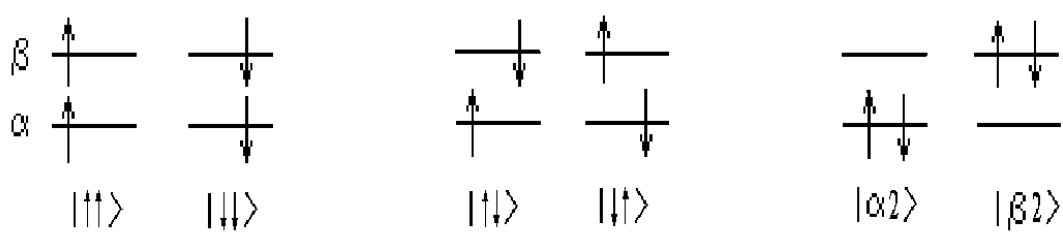
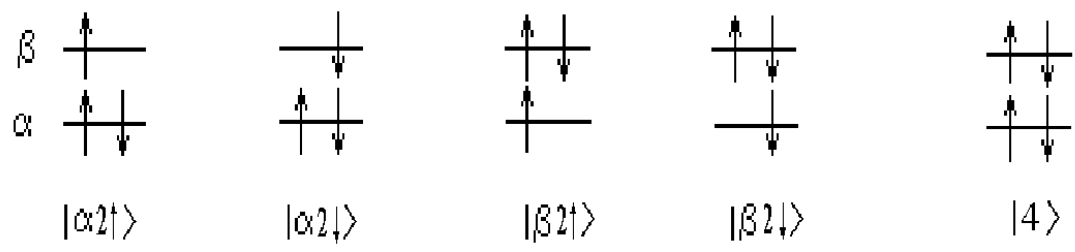
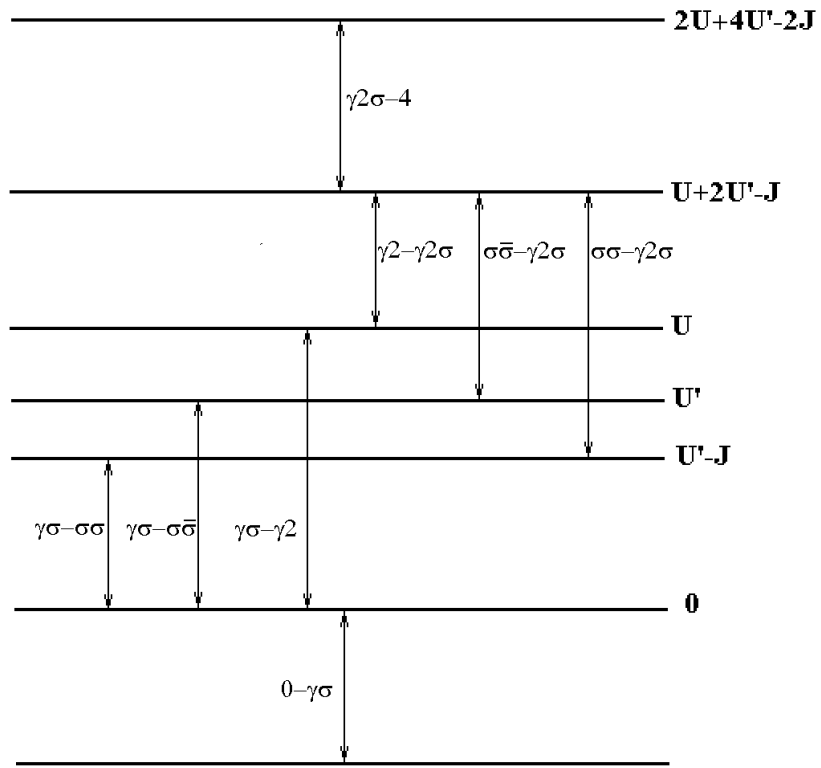
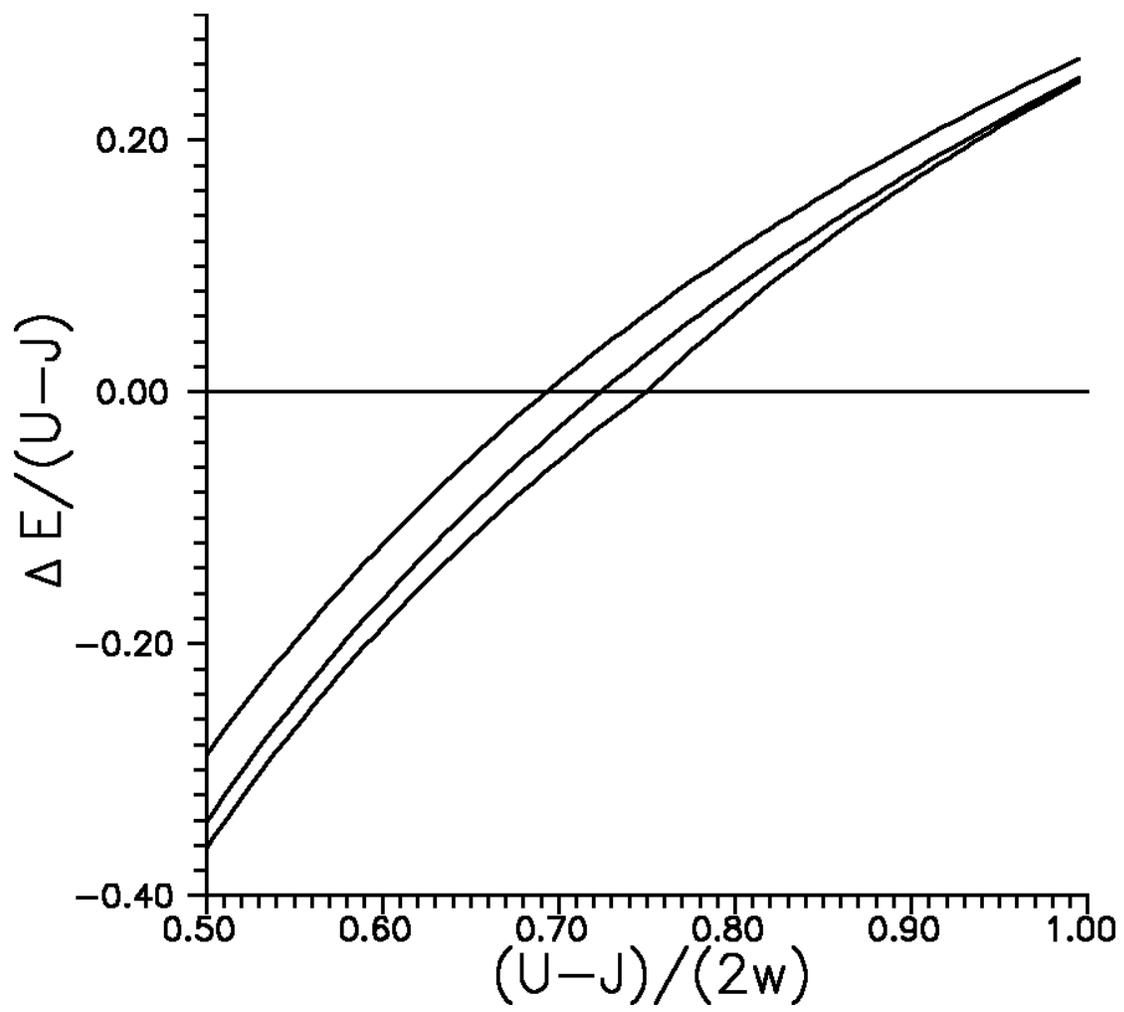
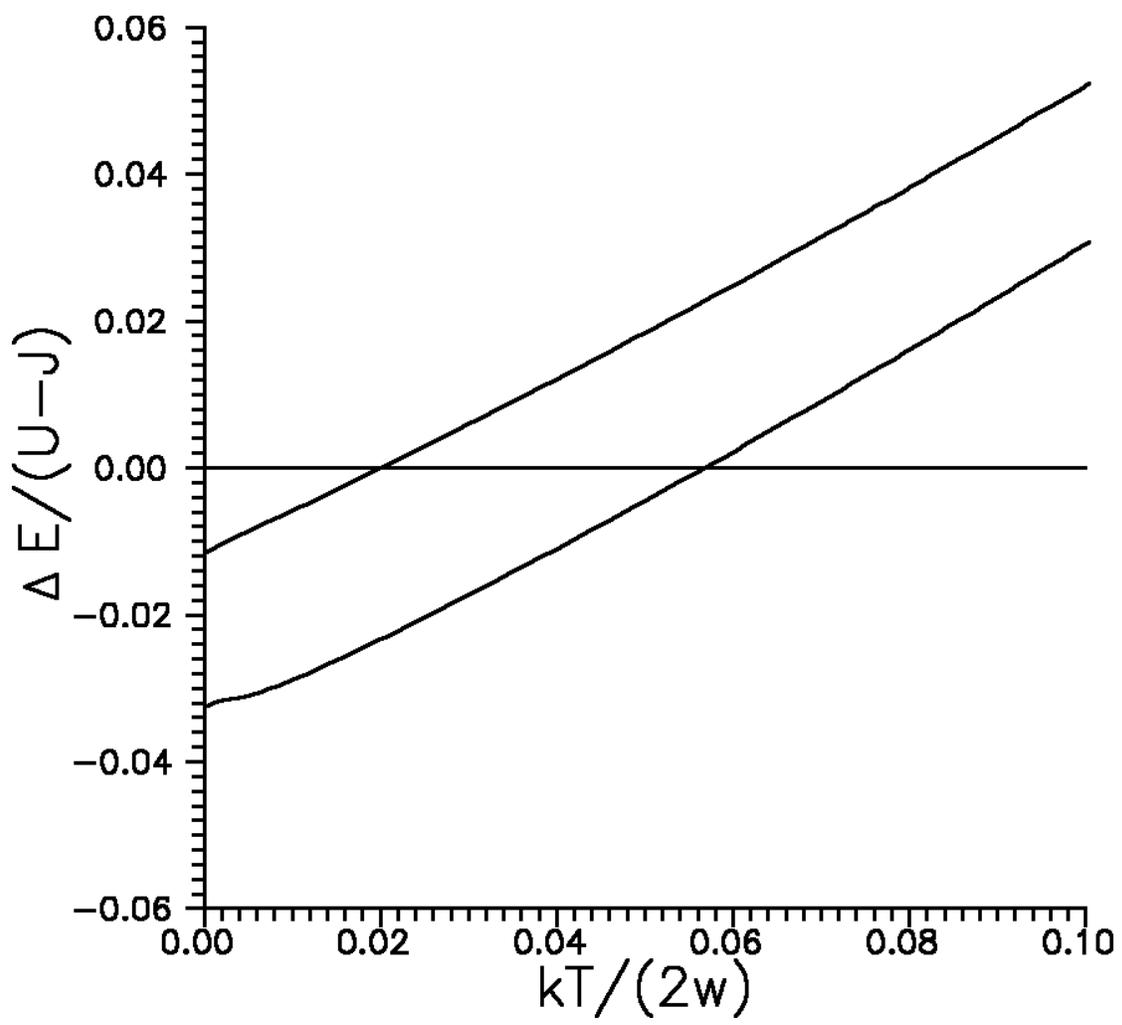
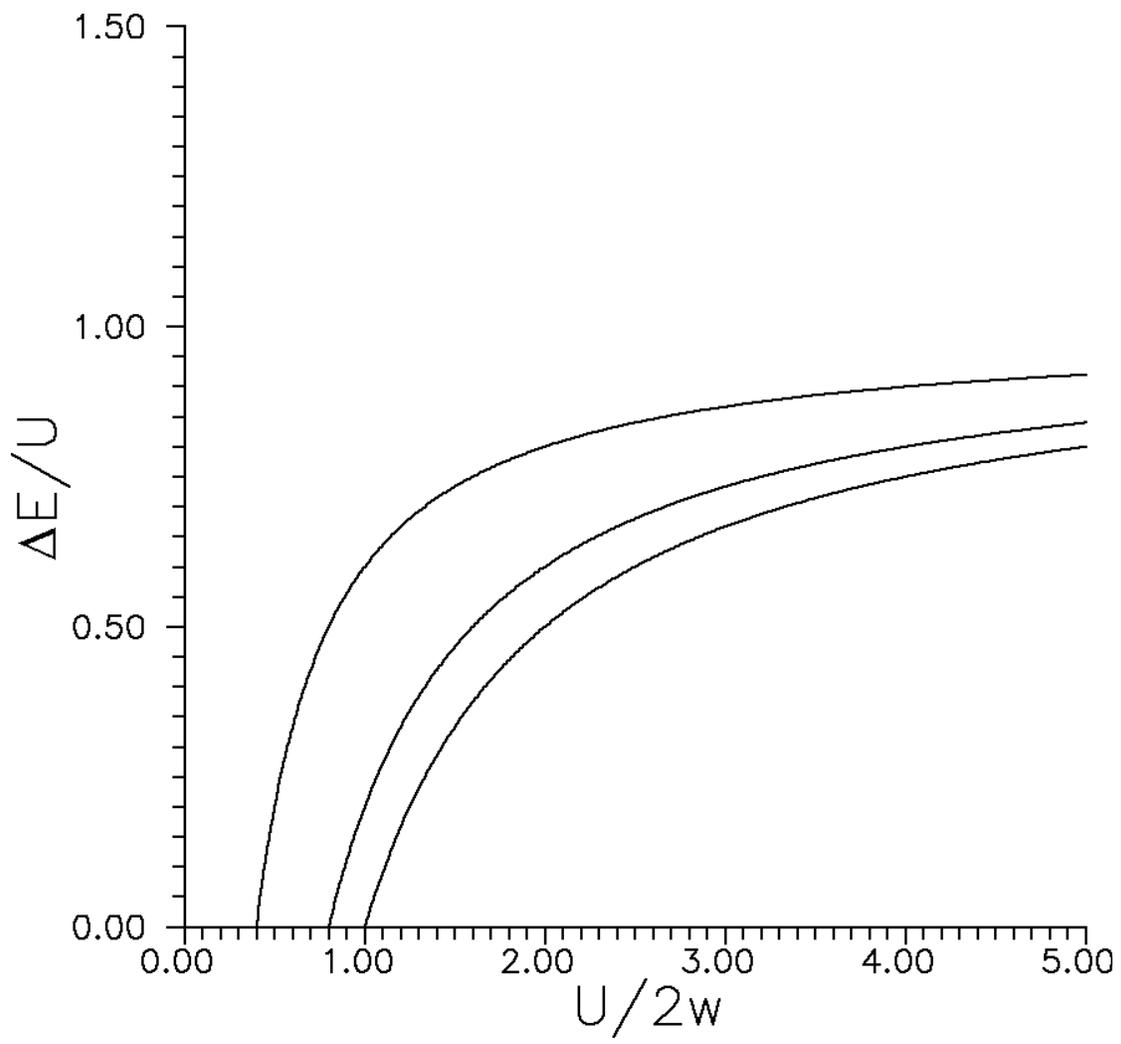
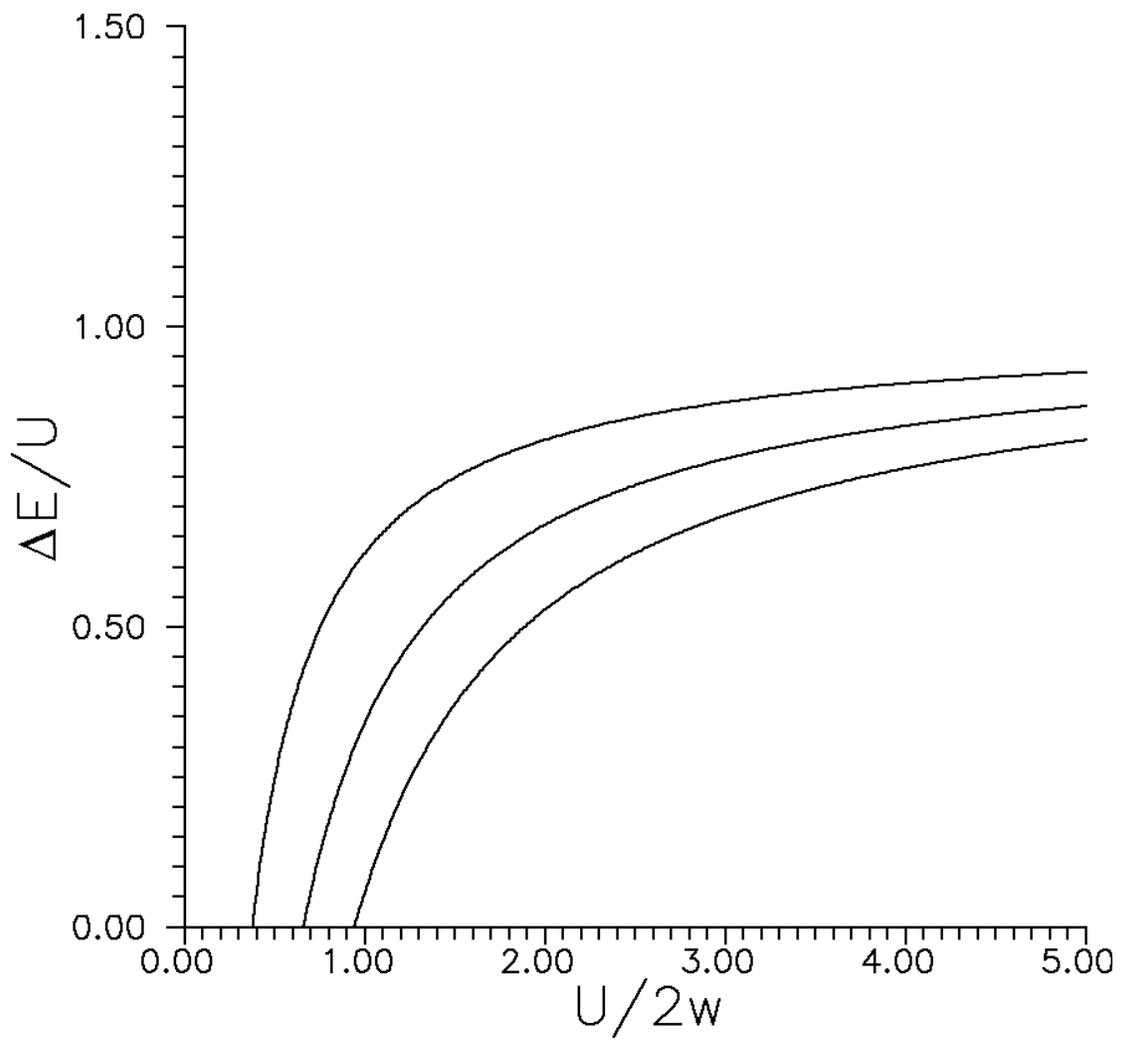
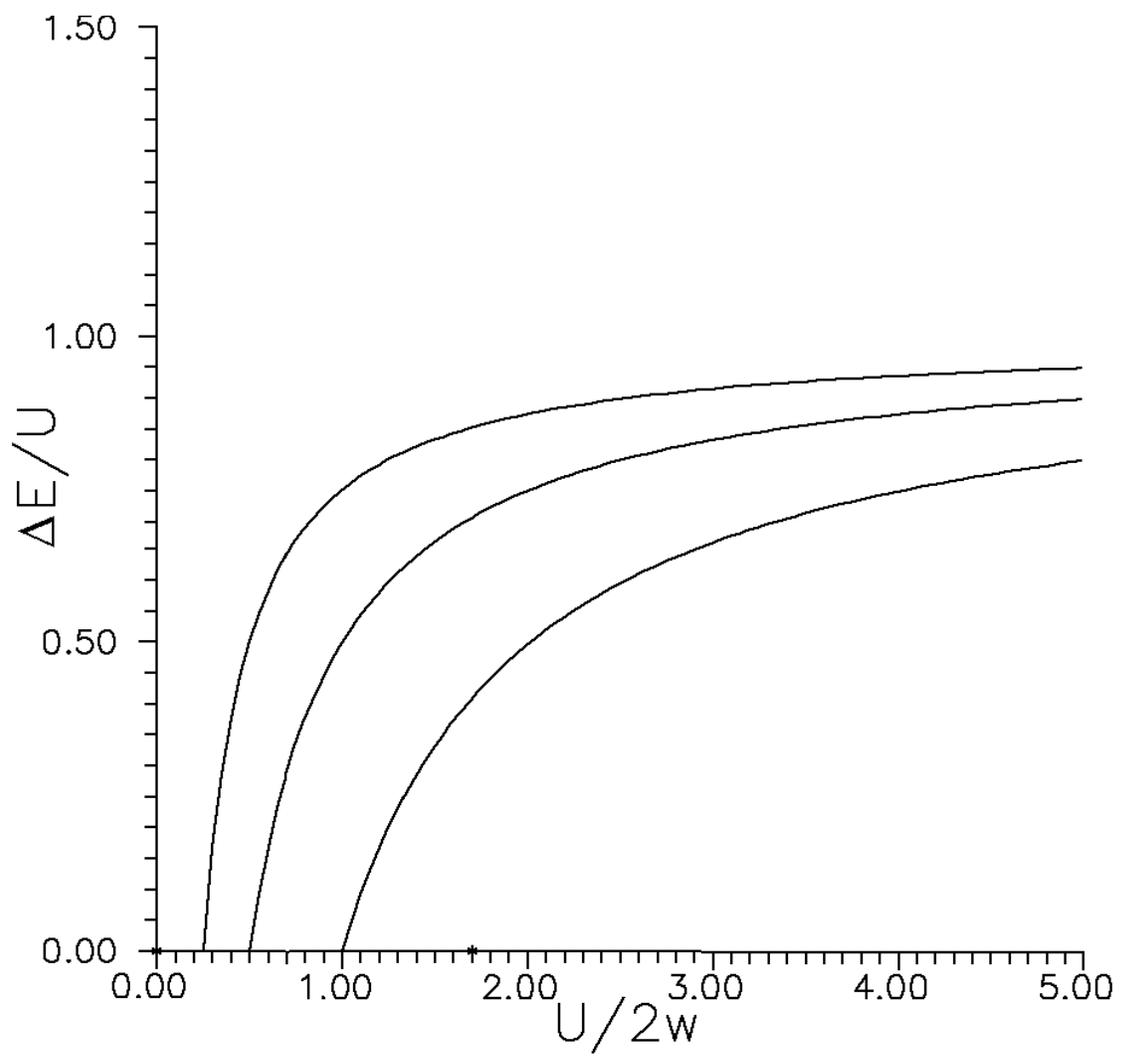
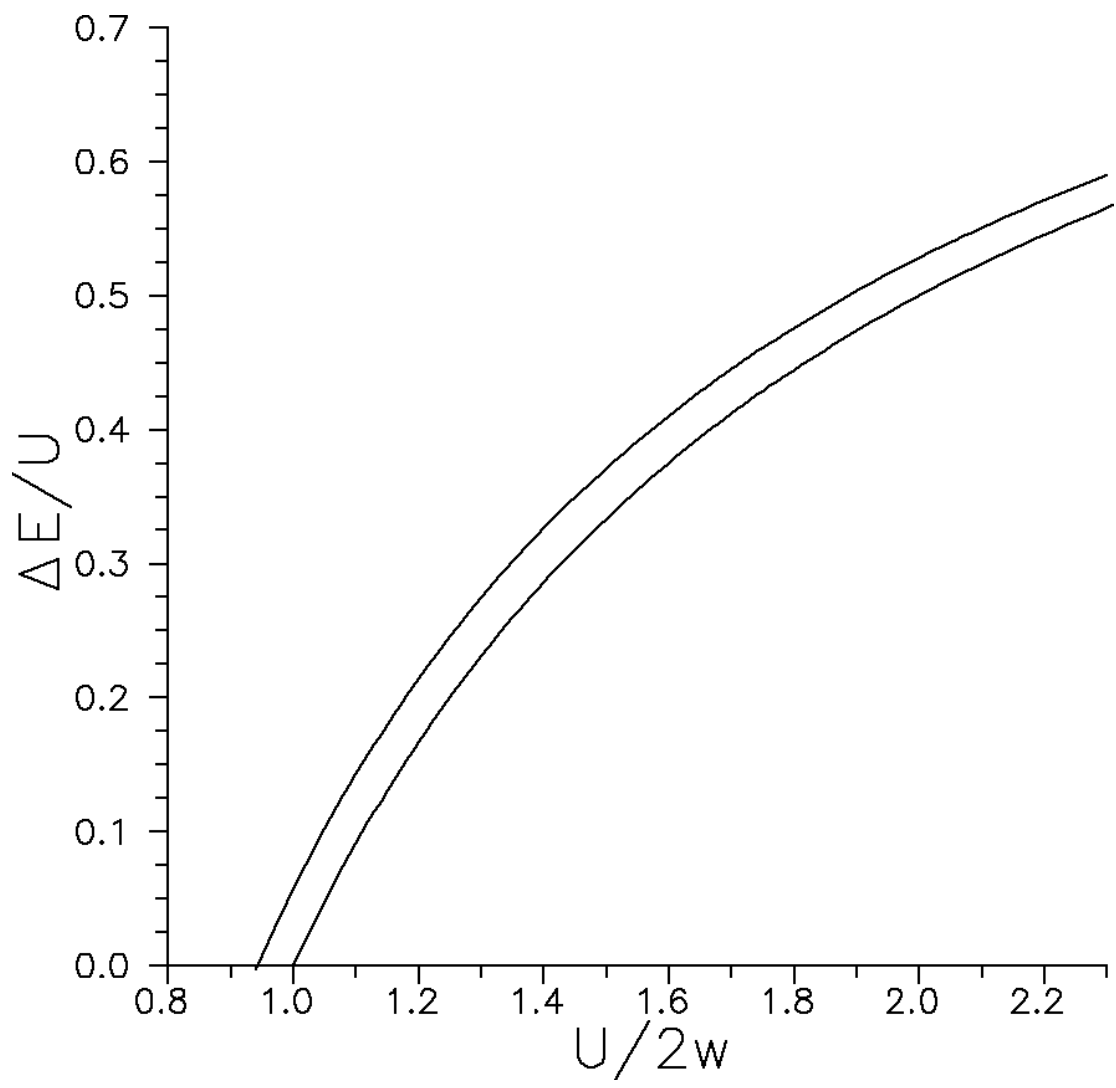
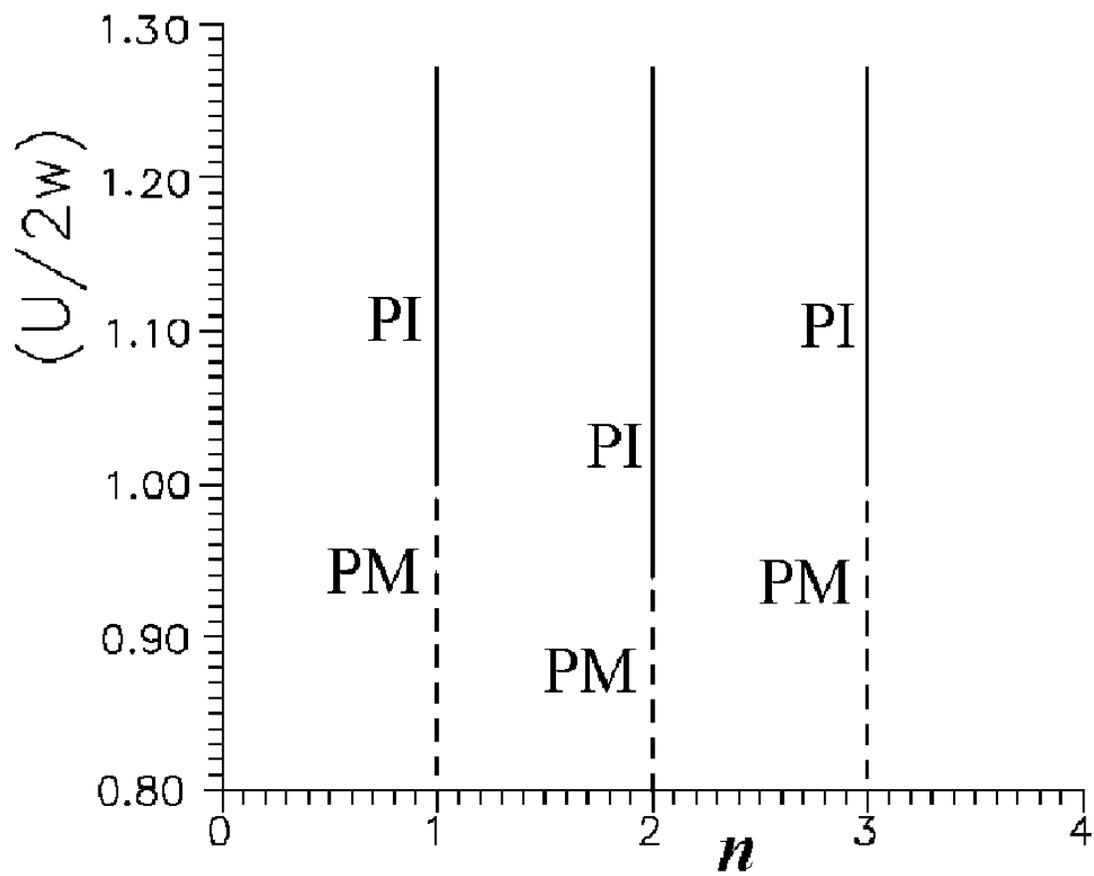
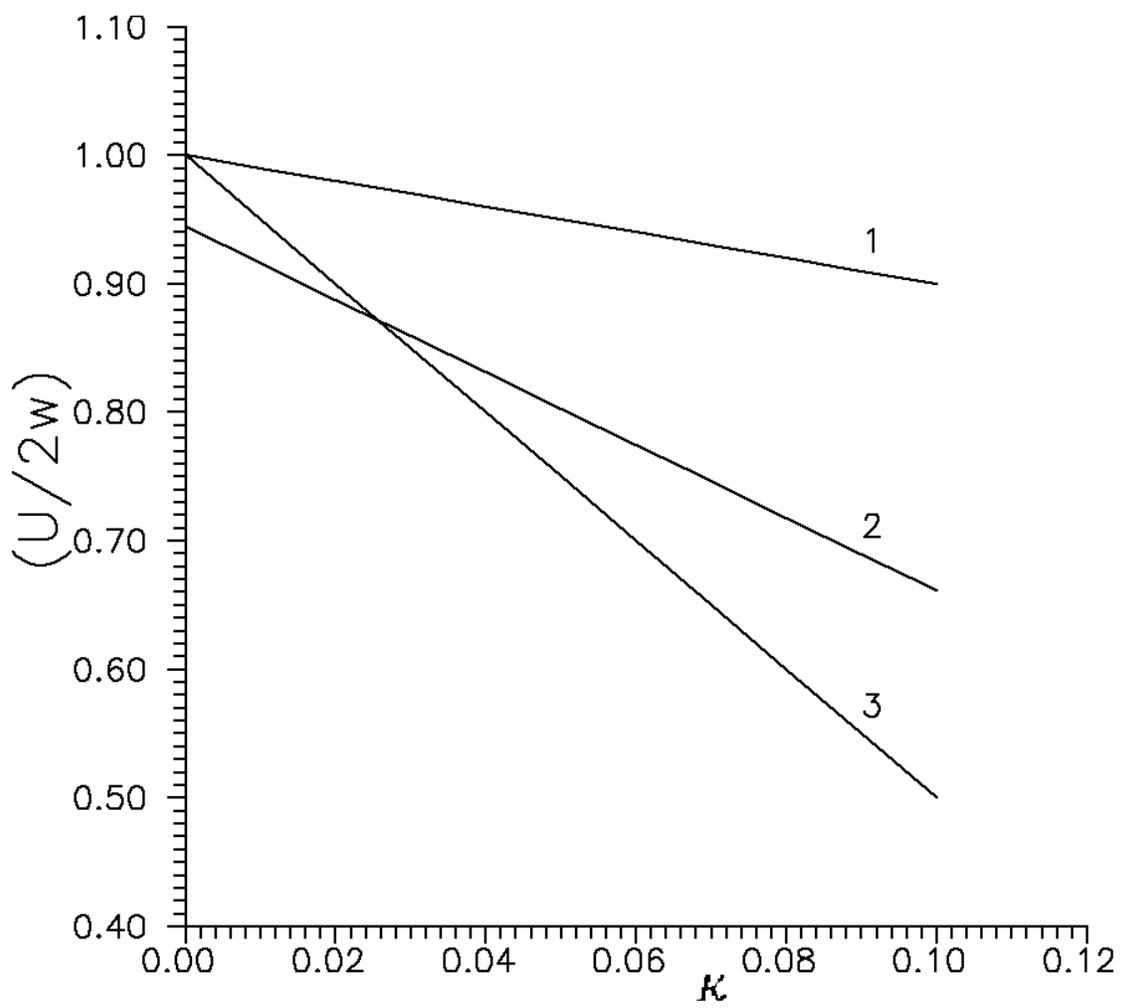
\end{document}